\documentclass[twocolumn,aps,prd,preprintnumbers,showpacs,superscriptaddress,nofootinbib,amsmath,amssymb,floats,floatfix,showkeys,notitlepage]{revtex4}

\usepackage{lipsum}
\usepackage{graphicx}
\usepackage{subfigure}
\usepackage{palatino}
\usepackage{changes}
\usepackage{hyperref}
\hypersetup{colorlinks=true,linkcolor=blue,urlcolor=blue,citecolor=blue}
\usepackage[toc,page]{appendix}
\usepackage[normalem]{ulem}
\usepackage{adjustbox}
\usepackage{latexsym}
\usepackage{amsmath}
\usepackage{amssymb}
\usepackage{amsfonts}
\usepackage{times}
\usepackage{dcolumn}
\usepackage{bm}
\usepackage{tikz}
\usepackage{bigints}
\usepackage{array,tabularx,multirow,booktabs}
\usepackage{tabularx}
\usepackage{multirow}
\usepackage[tracking=true]{microtype}
\SetTracking{}{500}
\SetTracking{encoding={*}, shape=sc}{40}
\UseRawInputEncoding %for inputenc error%
\allowdisplaybreaks

\begin{document} \sloppy
\title{Dark matter effect on the weak deflection angle by black holes at the center of Milky Way and M87 galaxies}
\author{Reggie C. Pantig}
\email{reggie.pantig@dlsu.edu.ph}
\affiliation{Physics Department, De La Salle University, 2401 Taft Avenue, Manila, 1004 Philippines}
\author{Ali \"Ovg\"un}
\email{ali.ovgun@emu.edu.tr}
\homepage{https://www.aovgun.com}
\affiliation{Physics Department, Eastern Mediterranean University, Famagusta, 99628 North Cyprus via Mersin 10, Turkey}

\begin{abstract}
In this paper, we investigated the effect of dark matter on the weak deflection angle by black holes at the galactic center. We consider three known dark matter density profiles such as the Cold Dark Matter, Scalar Field Dark Matter, and the Universal Rotation Curve from the Burkert profile. To achieve this goal, we used how the positional angles are measured by the Ishihara et al. method based on the Gauss-Bonnet theorem on the optical metric. With the help of the non-asymptotic form of the Gauss-Bonnet theorem, the longitudinal angle difference is also calculated. First, we find the emergence of apparent divergent terms on the said profiles, which indicates that the spacetime describing the black hole-dark matter combination is non-asymptotic. We showed that these apparent divergent terms vanish when the distance of the source and receiver are astronomically distant from the black hole. Using the current observational data in the Milky Way and M87 galaxies, we find interesting behaviors of how the weak deflection angle varies with the impact parameter, which gives us some hint on how dark matter interacts with the null particles for each dark matter density profile. We conclude that since these deviations are evident near the dark matter core radius, the weak deflection angle offers a better alternative for dark matter detection than using the deviation from the black hole shadow. With the dark matter profiles explored in this study, we find that the variation of the values for weak deflection angle strongly depends on the dark matter mass on a particular profile.
\end{abstract}



\maketitle

\section{Introduction} \label{int}
Several decades ago, the existence of a black hole was a mystery and only was explored as a mathematical construct \cite{Schwarzschild_1916}. Until recently, the Event Horizon Telescope collaboration unveiled the first image of the shadow of a black hole in the electromagnetic regime \cite{Event2021a,Event2021b,Collaboration_2019}, which again confirmed the correctness of Einstein's General Theory of Relativity \cite{Einstein_1916} as a model for compact objects with an extreme gravitational field. With the only confirmation of black hole's physical existence, one can not underestimate the progress of theoretical research on the search for the most realistic model of a black hole, as well as its dynamical interactions to any astrophysical environments \cite{Ozel:2021ayr,Vagnozzi:2020quf,Allahyari:2019jqz,Vagnozzi:2019apd,Guerrero:2021ues,Berti:2015itd,Cunha:2018acu,Barausse:2020rsu,Cunha:2017qtt,Junior:2021dyw,Cunha:2019hzj,Abdujabbarov:2017pfw,Narzilloev:2021jtg}.

There are recent studies on this specific direction, using perhaps the most important yet mysterious astrophysical environment - the dark matter. Dark matter constitutes about $85\%$ of the total mass of the Universe \cite{Jarosik2011} and is used to explain the strange behavior of stars and galaxy dynamics. At this time of writing, dark matter particles called Weakly Interacting Massive Particles (WIMPS) remain elusive to Earth-based direct detection experiments. Although there was some positive result reported \cite{Bernabei2008,Bernabei2013,Bernabei2018}, it was later criticized due to the null results from other improved direct-detection experiments \cite{Angholer2016,Amole2017,Akerib2017}. Thus, from the theoretical perspective at least, one must find an alternative for dark matter detection. Recently, it was proposed that the Earth's crust itself contains millions of years of data and can act as a huge dark matter detector \cite{Baum2020}. Meanwhile, can we also consider an extreme object such as a black hole to detect imprints of dark matter? Numerous research studies recently appeared to explore such a possibility. There are black hole models that came from the solution of the Einstein field equation that includes dark matter. See for example Ref. \cite{Hou_2018b} where the authors considered a perfect fluid dark matter. Dark matter toy models are also considered in studying its effect to the shadow \cite{Konoplya_2019,Pantig2020b}, weak deflection angle \cite{Pantig:2020odu,Pantig2022}, and the intensity of electromagnetic flux radiation \cite{Saurabh2021}. Until recently, a method was formulated to extract a particular black hole metric combined with some known dark matter profiles, thus, modeling a realistic scenario of a supermassive black hole at the heart of a galaxy surrounded with dark matter \cite{Xu_2018}. 

Gibbons and Werner showed a new geometrical technique to calculate the weak deflection angle using the Gauss-Bonnet theorem (GBT) on the optical geometries for asymptotically flat spacetimes \cite{Gibbons_2008}. In this method, one can solve the integral of GBT
in an infinite domain bounded by the light ray.
Then, Werner extended this method to stationary spacetimes by employing the
Finsler-Randers type optical geometry with Nazim's osculating Riemannian manifolds \cite{Werner_2012}. Afterward, Ishihara et al. extended this method for finite-distances (huge impact parameter) instead of using the asymptotic receiver and source \cite{Ishihara_2016,Ishihara:2016sfv}. Next, T. Ono et al. applied the finite-distances method to the axisymmetric spacetimes \cite{Ono:2017pie}.
Crisnejo and Gallo \cite{Crisnejo:2018uyn} used the GBT to obtain the gravitational deflections of light in a plasma medium. Recently, Li et al. studied the finite-distance effects on weak deflection angle by using massive particles and Jacobi-Maupertuis Randers-Finsler metric within GBT \cite{Li:2020dln,Li:2020wvn}. For
more recent works involving non-asymptotic spacetimes, finite distance, and the use of GBT on exotic and dark matter, one can see \cite{Werner:2012rc,Gibbons:2008zi,Jusufi:2017mav,Jusufi:2017lsl,Ovgun:2018fnk,Ono:2018ybw,Jusufi:2017uhh,Javed:2019qyg,Arakida:2017hrm,Ovgun:2019wej,Gibbons:2015qja,Ovgun:2018prw,Ono:2018jrv,Ovgun:2018oxk,Javed:2019rrg,Crisnejo:2018ppm,Javed:2019ynm,Jusufi:2017vew,Belhaj:2020rdb,Ono:2019hkw,Ovgun:2020gjz,Javed:2019jag,Javed:2020lsg,Takizawa:2020egm,Javed:2020fli,Ovgun2020,Okyay:2021nnh,Javed:2020frq,Takizawa:2020dja,Fu:2021akc,Arakida:2020xil,Kumaran:2021rgj,Zhang:2021ygh}.

A common result emerges from the aforementioned studies: dark matter is still very difficult to detect through the deviation arising from its effect, say, on the shadow radius of a black hole. In Ref. \cite{Hou_2018a}, it was stated that the dark matter effect on the shadow only occurs when its mass $k$ is around $10^7$ orders of magnitude, which means the dark matter distribution must be concentrated near the black hole and comparable to the black hole's mass. The study also suggested more improvement on the current resolution capabilities of modern telescopes. The same conclusion was found in Ref. \cite{Xu_2018,Konoplya_2019,Jusufi_2019,Pantig2020b}. Even with the consideration of the baryonic matter or the available observational data for the dark matter spike density, little and indistinguishable effect was seen in the black hole shadow \cite{Nampalliwar2021}. 
In this work, we will calculate the weak deflection angle by black holes lurking at the center of galaxies and determine if we can use such a phenomenon for better dark matter detection. In particular, we will derive the weak deflection angle for a black hole surrounded by known dark matter profiles such as the Cold Dark Matter (CDM) \cite{Navarro:1995iw,Navarro:1996gj,Dubinski:1991bm,Xu_2018}, Scalar Field Dark Matter (SFDM) \cite{Spergel:1999mh,Boehmer:2007um,Hou_2018a}, and Universal Rotation Curve (URC) \cite{Persic:1995ru}. The metric of these black holes was derived using the formalism pioneered by Xu et al. in Ref. \cite{Xu_2018}. The derived spacetime metrics are usually formidable in their exact form but these still satisfy the weak and strong energy conditions which indicate their physicality. See for example \cite{Xu2021a,Xu2021b}. With Xu et al.'s formalism, these black hole-dark matter metrics are rather new and to the best of our knowledge, there are no existing studies on calculating its weak deflection angle. There are existing studies, however, about other black hole-dark matter models and the calculation of their weak deflection angle and gravitational lensing effects \cite{Kaiser1993,Metcalf2001,Ovgun2019,Ovgun2020,Pantig:2020odu,Atamurutov2022,Pantig2022}.

The paper is organized as follows: Sect. \ref{sec1} introduces the Ishihara et al. method in calculating the weak deflection angle with the use of the positional and longitudinal angles. In the subsections that follow, the weak deflection angles of black holes with dark matter profiles are derived and discussed. In Sect. \ref{conc} we state concluding remarks and include research prospects. Throughout the paper, we used the natural units $G=c=1$ and the metric signature $(-,+,+,+)$.

\section{Weak Deflection Angle by Gauss-bonnet theorem and Ishihara-Li Method for finite distance} \label{sec1}
\begin{figure}[htpb]
    \centering
    \includegraphics[width=\linewidth]{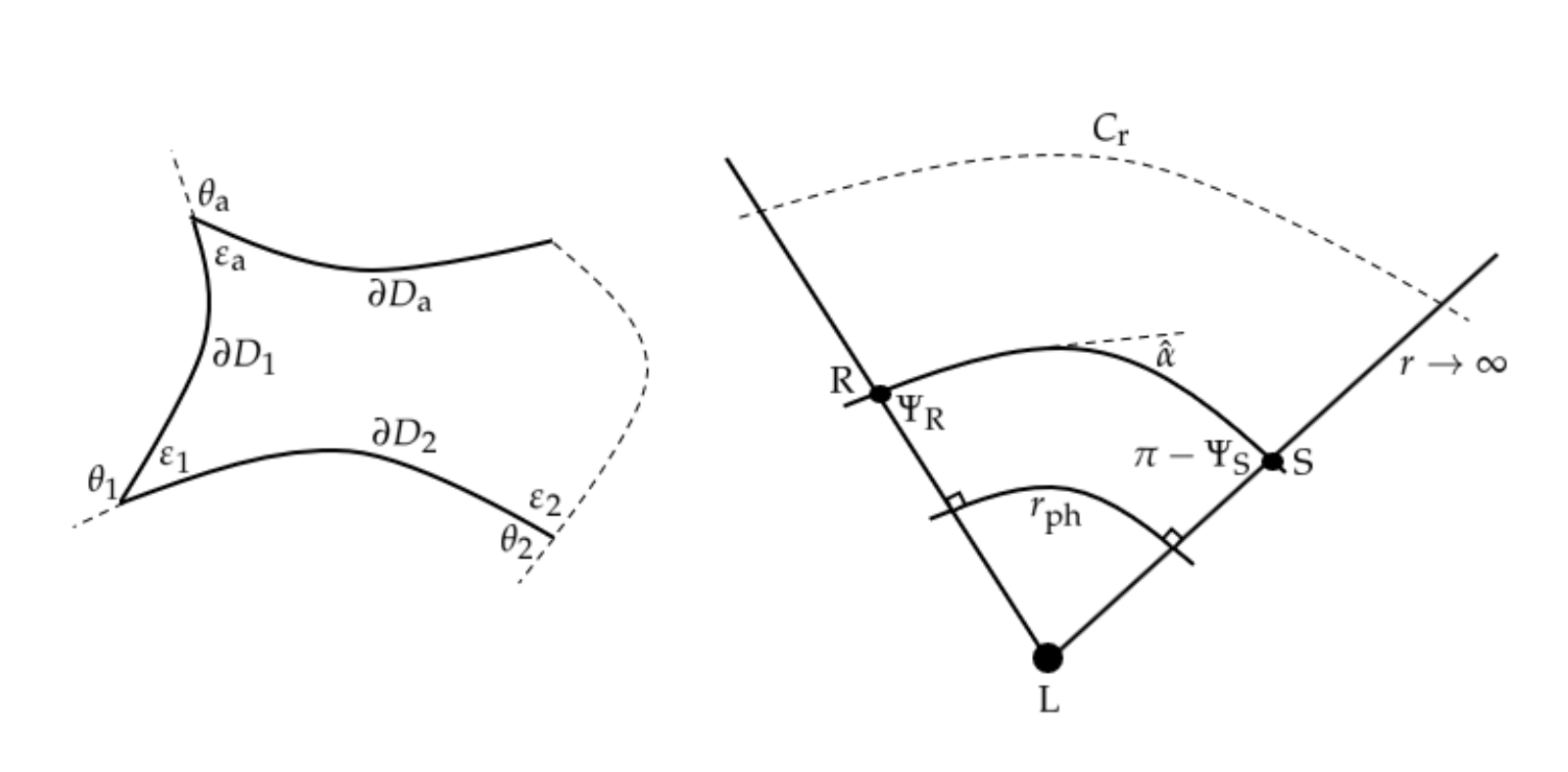}
    \caption{In the left is the schematic picture of a curved surface for Gauss-Bonnet theorem. In the right, the upper quadrilateral describes the domain for GBT as found by Ishihara et al. \cite{Ishihara_2016}, while the middle quadrilateral was used in Ref. \cite{Li:2020wvn} for non-asymptotic spacetimes.}
    \label{figGBT}
\end{figure}

Let us consider a non-rotating black hole whose spacetime in its vicinity is described as static and spherically symmetric:
\begin{equation} \label{e1}
    ds^{2} = -A(r) dt^{2} + B(r) dr^{2} + C(r) d\theta ^{2} +D(r) d\phi^{2},
\end{equation}
where $B(r)=A(r)^{-1}$, $C(r)=r^2$, and $D(r)=r^2\sin^2\theta$. In studying the weak deflection angle, light rays are important and they satisfy the null condition that $ds^2 = 0$. The optical metric can then be obtained via $dt=\sqrt{\gamma_{ij}dx^i dx^j}$, where $\gamma_{ij}$ is the spacial curve that runs from $1$ to $3$. With the optical metric, one can use it alongside with the Gauss-Bonnet theorem \cite{Carmo2016,Klingenberg2013} which states that
\begin{equation} \label{GBT}
    \iint_DKdS+\sum\limits_{a=1}^N \int_{\partial D_{a}} \kappa_{g} d\ell+ \sum\limits_{a=1}^N \theta_{a} = 2\pi
\end{equation}
to study deflection angles. Here, $D$ is any freely orientable 2D curved surface described by the Gaussian curvature $K$, and $dS$ is the its area element. The boundaries of $D$ are denoted by $\partial D_{\text{a}}$ (a=$1,2,..,N$) with the geodesic $\kappa_{\text{g}}$ integrated over the line element $d\ell$. Furthermore, $\theta_\text{a}$ and $\varepsilon_{\text{a}}$ are the jump and interior angles respectively. See Fig. \ref{figGBT}.

When the GBT is applied to a non-rotating black hole where the spacetime metric is asymptotically flat, Ishihara et al. \cite{Ishihara_2016} have shown that the weak deflection angle can be found by using the formula
\begin{equation} \label{e2}
    \hat{\alpha}=\phi_{\text{RS}}+\Psi_{\text{R}}-\Psi_{\text{S}} = -\iint_{_{\text{R}}^{\infty }\square _{\text{S}}^{\infty}}KdS.
\end{equation}

Here, $\Psi_{R}$ and $\Psi_{S}$ are the angles at the location of the receiver R and the source S respectively, and $\phi_{\text{RS}}$ is the coordinate separation angle between the receiver and the source, which is equal to the difference between the longitudes $\phi_R$ and $\phi_S$. It is further shown in Eq. \eqref{e2} how these angles are related to the GBT where the Gaussian optical curvature $K = \frac{R_{r\phi r\phi }}{\gamma }$ is integrated over the quadrilateral $_{R}^{\infty }\square _{S}^{\infty}$. Here, $\gamma$ is defined as the determinant of the optical metric $\gamma_{ij}$. Ishihara et al. \cite{Ishihara_2016} also proved that when the finite distances of the source and the receiver are considered, the LHS of the Eq. \eqref{e2} is equivalent to
\begin{equation} \label{e3}
    \hat{\alpha}=\int_{u_{R}}^{u_{o}}\frac{1}{\sqrt{F(u)}}du+\int_{u_{S}}^{u_{o}}\frac{1}{\sqrt{F(u)}}du+\Psi_{\text{R}}-\Psi_{\text{S}},
\end{equation}
where it is clear how $\phi_{\text{RS}}$ should be calculated. Here, $F(u)$ is the orbit equation expressed in terms of the inverse $r$-coordinate (ie. $r=1/u$):
\begin{equation} \label{e4}
    \left(\frac{du}{d\phi}\right)^2 \equiv F(u)=\frac{u^4}{b^2}\frac{C(u)(C(u)-A(u)b^2)}{A(u)B(u)}.
\end{equation}
The upper integration limit $u_o$ in Eq. \eqref{e3} is the iterative solution to Eq. \eqref{e4} while the boundary condition $\frac{du}{d\phi}\big|_{\phi=\frac{\pi}{2}}=0$ is imposed. Eq. \eqref{e3} is an elegant equation, handling non-asymptotically flat spacetimes where Eq. \eqref{e2} fails to work. An example would be those metrics that involve the cosmological constant, or those that have $r$ and $r^2$ terms in the metric function.

Recently, the study conducted in Ref. \cite{Li:2020wvn} also managed to use the GBT to calculate the weak deflection angle of non-asymptotically flat spacetimes. To do so, they used the middle quadrilateral in Fig. \ref{figGBT} which involves the photon radius, or in general, the circular orbit of a time-like particle. In their formalism, the weak deflection angle can be calculated using
\begin{equation} \label{en6}
    \hat{\alpha}=\phi_{\text{RS}}+\Psi_{\text{R}}-\Psi_{\text{S}} = \iint_{_{r_{\text{co}}}^{R }\square _{r_{\text{co}}}^{S}}KdS + \phi_{\text{RS}},
\end{equation}
where $r_{\text{co}}$ is replaced by $r_{\text{ph}}$ for photon deflection angle. The interested reader is invited to look on Ref. \cite{Li:2020wvn} for the complete treatment of their method and applications.

Now based on observing Eqs. \eqref{e2} and \eqref{en6}, it occurs that it is easier to calculate the weak deflection angle by using the original definition of $\hat{\alpha}=\phi_{\text{RS}}+\Psi_{\text{R}}-\Psi_{\text{S}}$ since it avoids the task of integrating a particular expression for the Gaussian optical curvature. As we know, metric functions that contain simple expressions are easy to calculate using the GBT, especially if these functions are derived as a solution to the Einstein field equation. However, there exist some metric functions that are complicated enough that integrating their Gaussian optical curvature, or by integrating the inverse-root of the orbit equation, gives no analytical expression or very complicated results. Examples of these metrics are the ones derived using Xu et al. \cite{Xu_2018} formalism that involves a black hole surrounded by dark matter described by a density profile. See Refs. \cite{Xu2021a,Xu2021b} for example, and the one recently derived by Authors \cite{Jusufi_2019,Nampalliwar2021,Hou_2018a}.

In this paper, since we are interested in deriving the weak deflection angles by black holes at the center of a galaxy surrounded by dark matter, we will avoid such integrals in the expressions in Eqs. \eqref{e2} and \eqref{en6}. Instead, we will focus on the calculation of the positional angles $\Psi$ and longitudinal angle $\phi$. We can calculate $\Psi$ by going back to Eq. \eqref{e2}, where the angles $\Psi_{R}$ and $\Psi_{S}$, $\cos \Psi\equiv\gamma_{ij}e^iR^j$ can be defined using the inner product of the unit basis vector $e^i$ along the equatorial plane, and the unit radial vector $R^i$ with relative to the lensing object \cite{Ishihara_2016}: ie.
\begin{align} \label{e5}
    e^i&=\left(\frac{dr}{dt},0,\frac{d\phi}{dt}\right)=\frac{d\phi}{dt}\left(\frac{dr}{d\phi},0,1\right) \nonumber \\
    R^i&=\left(\frac{1}{\sqrt{\gamma_{rr}}},0,0\right).
\end{align}
By using the orbit equation $F(r)$, $\cos\Psi$ can be recasted as
\begin{equation} \label{e6}
    \sin\Psi=\sqrt{\frac{A(r)}{C(r)}}b
\end{equation}
which is more convenient to calculate than $\cos\Psi$. Finally, for the calculation of the longitudinal angle $\phi$, it is done by iteratively solving the orbit equation in Eq. \eqref{e4} \cite{Ono:2019hkw}.

\subsection{Effect of the CDM profile on weak deflection angle by black holes} \label{sec2}
The Cold Dark Matter density profile is one of the most well-known profile that is consistent with astronomical observations in the large scale \cite{Navarro:1995iw,Navarro:1996gj,Dubinski:1991bm}. Although the physical nature of dark matter is unknown, dark matter particles are modeled with non-relativistic motion. The CDM density profile is expressed
\begin{equation} \label{e7}
    \rho = \frac{\rho_\text{c}}{\frac{r}{r_\text{c}}\left(1+\frac{r}{r_\text{c}}\right)^2},
\end{equation}
where $\rho$ is the Universe's density at the time of dark matter collapse, while $\rho_\text{c}$ and $r_\text{c}$ are the core density and radius respectively. It is shown in \cite{Xu_2018} how we can obtain the black hole metric function with the CDM profile associated with Eq.\eqref{e7}. First, the mass profile for the dark matter halo is calculated as
\begin{equation} \label{e8}
M_{\text{DM}}(r)=4 \pi \int_{0}^{r} \rho\left(r^{\prime}\right) r^{\prime 2} d r^{\prime}.
\end{equation}
Then using the mass profile, the tangential velocity of test particle in dark matter halo is calculated easily by $v_{\text{tg}}^{2}(r)=M_{\text{DM}}(r) / r$. On the other hand, if the line element describing the dark matter halo is given by
\begin{equation} \label{e9}
    ds^{2} = -f(r) dt^{2} + f(r)^{-1} dr^{2} + r^2 d\theta ^{2} +r^2\sin^2\theta d\phi^{2},
\end{equation}
we can derive a rotational velocity of a test particle in static and spherical symmetric spacetime of the DM halo using the relation
\begin{equation} \label{e10}
v_{\text{tg}}^{2}(r)=\frac{r}{\sqrt{f(r)}} \frac{d \sqrt{f(r)}}{d r}=r \frac{d \ln (\sqrt{f(r)})}{d r}.
\end{equation}
Now, using the above relations for the rotation velocities, the metric function of the dark matter halo can be derived by using:
\begin{equation} \label{e11}
f(r)=\exp \left[2 \int \frac{v_{\text{tg}}^{2}(r)}{r} d r\right].
\end{equation}
From the Einstein field equation, the idea is to combine the dark matter profile and the Schwarzschild black hole to the energy-momentum tensor $T^{\mu}_{\nu}$. If the resulting metric is expressed by Eq. \eqref{e1}, then it is defined that $A(r)=f(r)+F(r)$ where $F(r)$ the metric coefficient that describes the black hole. For details, see Ref. \cite{Xu_2018}. One then finds that the metric function of the black hole in CDM profile as
\begin{equation} \label{e12}
    A(r)=\left(1+\frac{r}{r_\text{c}}\right)^{-\frac{8\pi k}{r}}-\frac{2m}{r},
\end{equation}
where $m$ is the mass of the black hole. Here, we denote $k$ in as the dark matter mass given by $k=\rho_\text{c} r_\text{c}^3$ for simplicity.

The orbit equation $F(u)$ can then be easily calculated using Eq. \eqref{e4} as
\begin{equation} \label{e13}
    F(u) = \frac{1}{b^{2}} - u^{2} + 2mu^{3} + 8\pi u^3 k \ln\left(1+\frac{1}{ur_\text{c}}\right),
\end{equation}
where the first three terms are the known orbit equation for the Schwarzschild case and the last term is the dark matter contribution. We differentiate again Eq. \eqref{e13} with respect to $\phi$ and obtain
\begin{equation}
    -\frac{d^{2}u}{d\phi^{2}}+4\pi ku^{2}\left[3\ln\left(1+\frac{1}{\kappa u}\right)-\frac{1}{\kappa u+1}\right]+3mu^{2}-u=0.
\end{equation}
Here, we think that $ku$ and $mu$ are so small, and this procedure in perturbation method allows us to solve the differential equation
\begin{equation}
    -\frac{d^{2}u}{d\phi^{2}}-u=0,
\end{equation}
which gives
\begin{equation}
    u_\text{o}=X\sin(\phi)+Y\sin(\phi).
\end{equation}
We use the boundary condition that $\frac{du}{d\phi}\big|_{\phi=\frac{\pi}{2}}=0$, and this implies that $X=1/b$. Thus,
\begin{equation} \label{e19n}
    u_\text{o}=\frac{\sin(\phi)}{b}.
\end{equation}
Beginning from Eq. \eqref{e19n}, we can proceed to solve Eq. \eqref{e13} by iteration resulting to
\begin{equation} \label{e14}
    u_o=\frac{\sin\phi}{b}+\frac{m\left(1+\cos^{2}\phi\right)}{b^{2}}+ \frac{4\pi k}{b^{2}}\ln\left(1+\frac{b}{r_\text{c}}\right).
\end{equation}
Using this, we can solve the angle $\phi$ directly. We then obtained
\begin{align} \label{e15}
    \phi_{\text{RS}}&=(\phi_{\text{RS}})_{\text{Schw}} \nonumber\\
    &+\frac{4\pi k}{b}\ln\left(1+\frac{b}{r_\text{c}}\right)\left[\frac{1}{\sqrt{1-b^{2}u_{\text{R}}^{2}}}+\frac{1}{\sqrt{1-b^{2}u_{\text{S}}^{2}}}\right] \nonumber\\
    &-4\pi bkm\left[\frac{u_{\text{R}}^{3}\ln\left(1+\frac{b}{r_\text{c}}\right)}{\left(1-b^{2}u_{\text{R}}^{2}\right)^{3/2}}+\frac{u_{\text{S}}^{3}\ln\left(1+\frac{b}{r_\text{c}}\right)}{\left(1-b^{2}u_{\text{S}}^{2}\right)^{3/2}}\right]\nonumber \\
    &+\mathcal{O}(m^2,k^2,m^2k^2),
\end{align}
where $u_\text{S}$ and $u_\text{R}$ are the inverse of the radial distance of the source and the receiver from the lensing object respectively. For simplicity, we also have written
\begin{align} \label{e16}
    (\phi_{\text{RS}})_{\text{Schw}}&=\pi-\arcsin bu_{\text{R}}-\arcsin bu_{\text{R}} \nonumber\\
    &-\frac{m}{b}\left[\frac{\left(b^{2}u_{\text{R}}^{2}-2\right)}{\sqrt{1-b^{2}u_{\text{R}}^{2}}}+\frac{\left(b^{2}u_{\text{S}}^{2}-2\right)}{\sqrt{1-b^{2}u_{\text{S}}^{2}}}\right].
\end{align}
Now, using Eq. \eqref{e6}, we have
\begin{align} \label{e17}
    \Psi_{\text{R}}-\Psi_{\text{S}}&=(\Psi_{\text{R}}-\Psi_{\text{S}})_{\text{Schw}} \nonumber\\
    &-4\pi bk\left[\frac{u_{\text{R}}^{2}\ln\left(1+\frac{1}{r_\text{c} u_{\text{R}}}\right)}{\sqrt{1-b^{2}u_{\text{R}}^{2}}}+\frac{u_{\text{S}}^{2}\ln\left(1+\frac{1}{r_\text{c} u_{\text{S}}}\right)}{\sqrt{1-b^{2}u_{\text{S}}^{2}}}\right] \nonumber\\
    &+4\pi bkm\Bigg[\frac{u_{\text{R}}^{3}\left(2b^{2}u_{\text{R}}^{2}-1\right)\ln\left(1+\frac{1}{r_\text{c} u_{\text{R}}}\right)}{\left(1-b^{2}u_{\text{R}}^{2}\right)^{3/2}} \nonumber\\
    &+\frac{u_{\text{S}}^{3}\left(2b^{2}u_{\text{S}}^{2}-1\right)\ln\left(1+\frac{1}{r_\text{c} u_{\text{S}}}\right)}{\left(1-b^{2}u_{\text{S}}^{2}\right)^{3/2}}\Bigg] \nonumber\\
    &+\mathcal{O}(m^2,k^2,m^2k^2),
\end{align}
where
\begin{align} \label{e18}
    (\Psi_{\text{R}}-\Psi_{\text{S}})_{\text{Schw}}&=-\pi+\arcsin bu_{\text{R}}+\arcsin bu_{\text{R}} \nonumber\\
    &-bm\left[\frac{u_{\text{R}}^{2}}{\sqrt{1-b^{2}u_{\text{R}}^{2}}}+\frac{u_{\text{S}}^{2}}{\sqrt{1-b^{2}u_{\text{S}}^{2}}}\right].
\end{align}
Combining the two previous equations above, we obtain the weak deflection angle with finite distance of the source and the receiver as
\begin{widetext}
\begin{align} \label{e19}
    \hat{\alpha}&=\frac{2m}{b}\left(\sqrt{1-b^{2}u_{\text{R}}^{2}}+\sqrt{1-b^{2}u_{\text{S}}^{2}}\right)+\frac{4\pi k}{b}\left\{ \frac{\left[b^{2}u_{\text{R}}^{2}\ln\left(1+\frac{1}{r_\text{c} u_{\text{R}}}\right)+\ln\left(1+\frac{b}{r_\text{c}}\right)\right]}{\sqrt{1-b^{2}u_{\text{R}}^{2}}}+\frac{\left[b^{2}u_{\text{S}}^{2}\ln\left(1+\frac{1}{r_\text{c} u_{\text{S}}}\right)+\ln\left(1+\frac{b}{r_\text{c}}\right)\right]}{\sqrt{1-b^{2}u_{\text{S}}^{2}}}\right\} \nonumber\\
    &-4\pi bkm\Bigg\{ \frac{u_{\text{R}}^{3}\left[\left(2b^{2}u_{\text{R}}^{2}-1\right)\ln\left(1+\frac{1}{r_\text{c} u_{\text{R}}}\right)-\frac{1}{2}\ln\left(1+\frac{b}{r_\text{c}}\right)\right]}{\sqrt{1-b^{2}u_{\text{R}}^{2}}}+\frac{u_{\text{S}}^{3}\left[\left(2b^{2}u_{\text{S}}^{2}-1\right)\ln\left(1+\frac{1}{r_\text{c} u_{\text{S}}}\right)-\frac{1}{2}\ln\left(1+\frac{b}{r_\text{c}}\right)\right]}{\sqrt{1-b^{2}u_{\text{S}}^{2}}}\Bigg\} \nonumber\\
    &+\mathcal{O}(m^2,k^2,m^2k^2).
\end{align}
\end{widetext}
We note that this leads to the weak deflection angle $\hat{\alpha}=4m/b$ in the Schwarzschild case when there is no dark matter mass ($k=0$) in the far approximation. We can also see in Eq. \eqref{e19} how the value of $u$ depends on the impact parameter $b$. For $\hat{\alpha}$ to have some physical significance, $u$ should not be any lower than $1/b$, which means that the position of the source and receiver should not be less than $b$.

In Eq. \eqref{e19}, the quantity $\ln(1+1/r_\text{c} u)$ is undefined when $u$ is exactly zero. However, if $u$ has a finite value no matter how near it is to zero, there is a certain finite value for $\ln(1+1/r_\text{c} u)$. Notice also that the $u^2$ and $u^3$ as scaling factor dominates the $\ln(1+1/r_\text{c} u)$ and these terms can be safely approximated to zero. Hence, assuming that $u_{\text{R}}=u_{\text{S}}$ and are very small, Eq. \eqref{e19} can still be approximated as
\begin{equation} \label{e20}
    \hat{\alpha}=\frac{4m}{b}+\frac{8\pi k}{b}\ln\left(1+\frac{b}{r_\text{c}}\right).
\end{equation}

Let us use the available data for Sgr. A* \cite{Hou_2018a} where the values of the dark matter core density and core radius are $\rho_c=1.936$x$10^7M_\odot\text{kpc}^{-3}$ and $r_\text{c} = 17.46\text{ kpc}$ respectively. The dark matter mass is then $k = 1.030$x$10^{11}M_\odot$ while the black hole mass at the center is $m = 4.30$x$10^{6}M_\odot$. It is useful to express $k$ in terms of the black hole mass unit and this gives $k=23953m$. Notice that if we compare $m$ to $r_\text{c}$, it is indeed that Eq. \eqref{e20} applies to the situation. We can still use this equation at the location where the "cusp" phenomenon begins to occur, which is $1$ kpc and below. \cite{deBlok_2010,Xu_2018}. We note, however, that one cannot use the equation when the impact parameter is very close to the black hole. Thus, a reasonable range for the impact parameter will be shown to demonstrate the effect of the dark matter of various profiles. Finally, Eq. \eqref{e19} is the general equation considering the finite distance of the source and the receiver and it is interesting to compare it to the approximated expression in Eq. \eqref{e20}.

\begin{figure}[htpb]
    \centering
    \includegraphics[width=\linewidth]{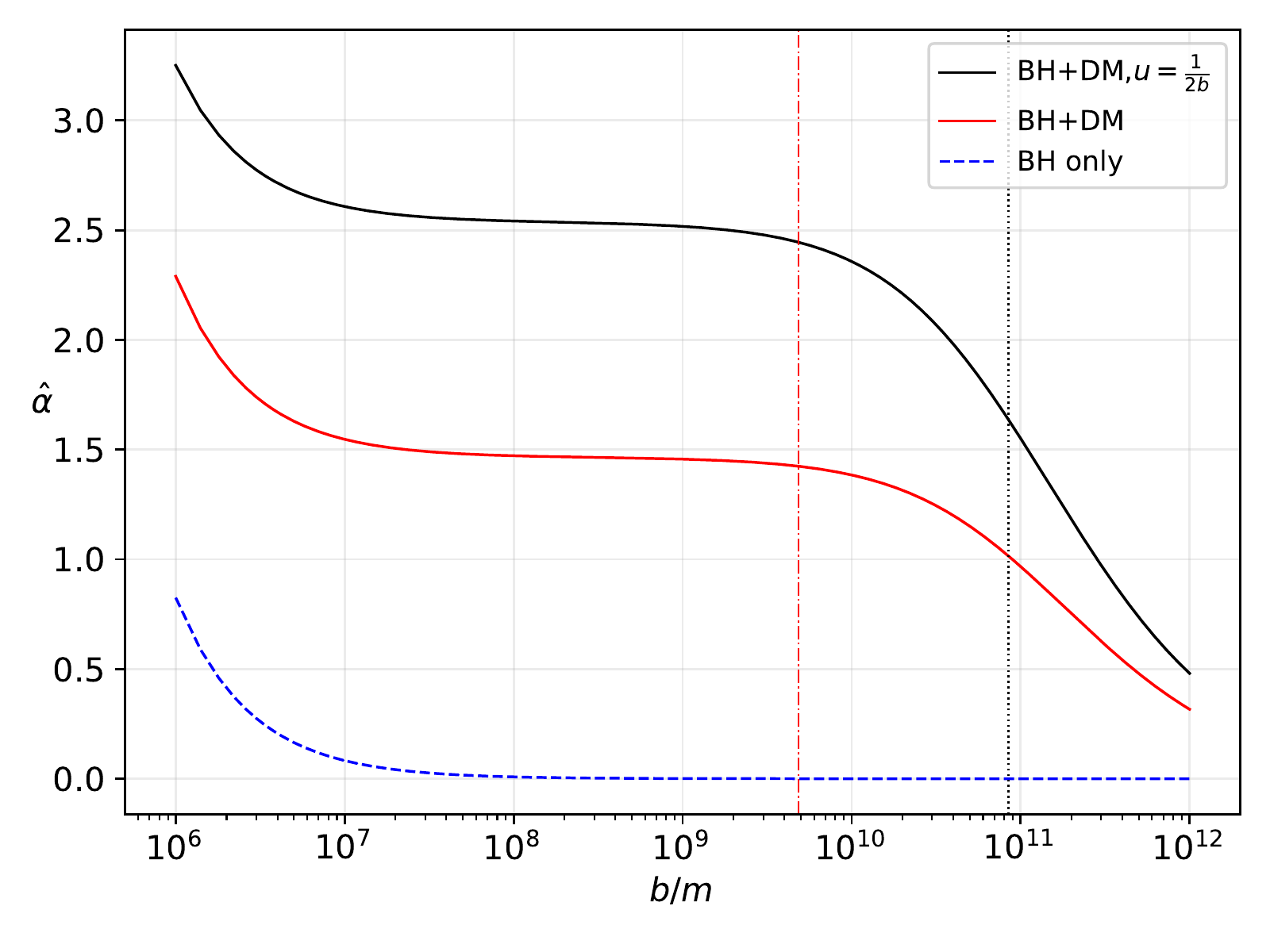}
    \caption{Curve of the weak deflection angle in the CDM profile for different values of the dimensionless impact parameter $b/m$ in Sgr. A*. Here, $\hat{\alpha}$ is in $\mu$as. The vertical dotted line is the value of the outermost core radius $r_\text{c}\sim8.484$x$10^{10}m$. The region below the dash-dotted vertical line is the cusp region ($\sim 4.86$x$10^9m$).}
    \label{fig1}
\end{figure}
\begin{figure}
    \centering
    \includegraphics[width=\linewidth]{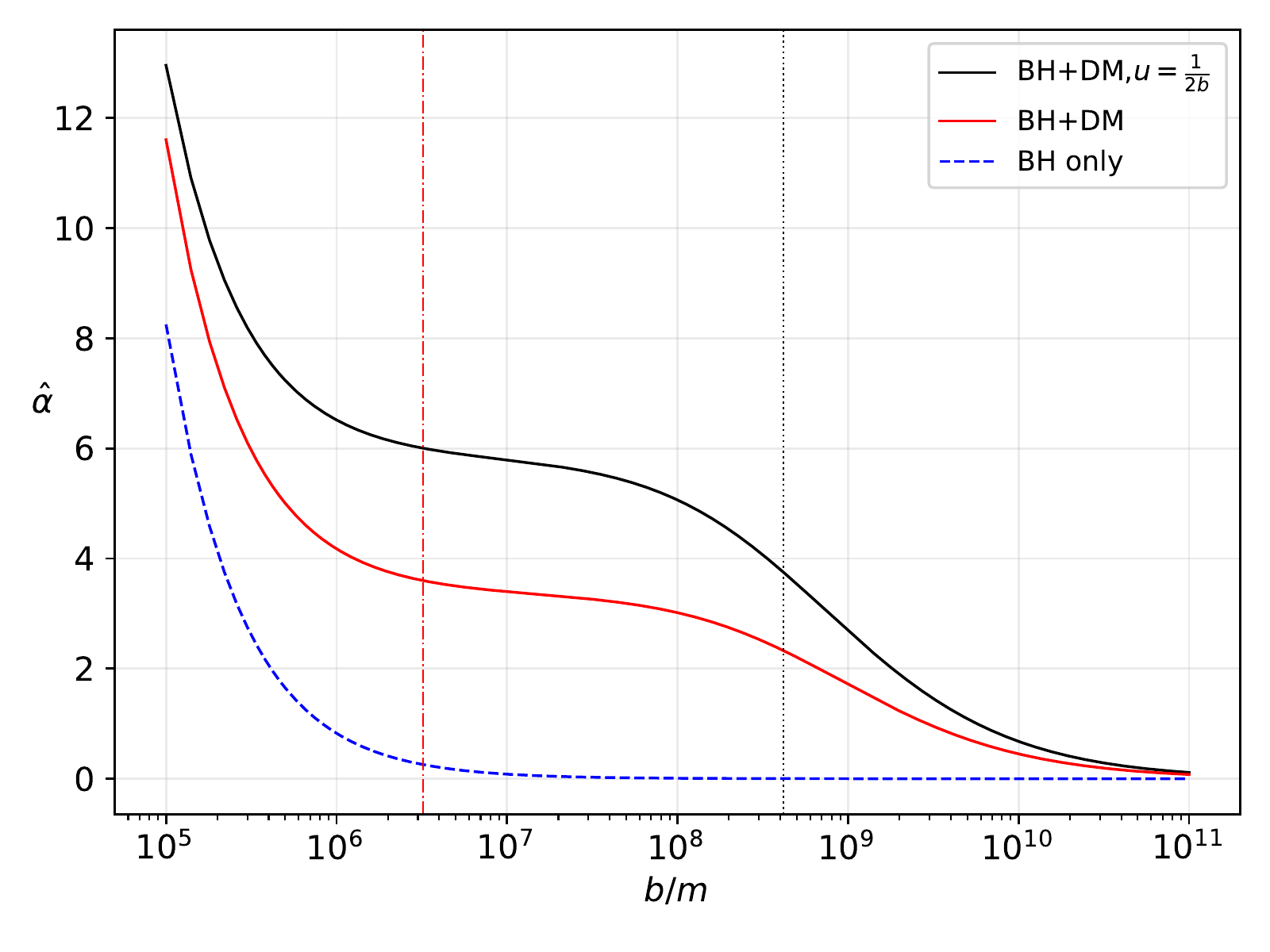}
    \caption{Curve of the weak deflection angle in the CDM profile for different values of the dimensionless impact parameter $b/m$ in M87. The vertical dotted line is the value of the outermost core radius $r_{\text{c}} \sim 4.179$x$10^{8}m$. The region below the dash-dotted vertical line is the cusp region ($\sim 3.214$x$10^6m$).}
    \label{fig1n}
\end{figure}
Fig. \ref{fig1} shows the plot of Eqs. \eqref{e19} and \eqref{e20} for Sgr. A*. The solid black line corresponds to the case where the positions of the source and receiver are twice the impact parameter. Meanwhile, the solid red line corresponds to the approximated case where $u$ is very small. For comparison, we also plotted the weak deflection angle $\hat{\alpha}$ by the black hole alone, which is so small considering the range for $b/m$ (blue dashed line). When the effect of the CDM profile is taken into account, we see some interesting deviations from the Schwarzschild case. Eq. \eqref{e20} tells us that when $b>>r_\text{c}$, the dark matter term dominates the standard value, thus the observed increase in $\hat{\alpha}$. As the impact parameter $b/m$ gets comparable with the core radius $r_\text{c}$, the rate at which $\hat{\alpha}$ changes is increased drastically. We see that the deviation is larger due to the effect of the finite distance of the source and the receiver. Interestingly, between the core radius and the cusp region, the rate at which $\hat{\alpha}$ increases levels off and then follows the same trend as the Schwarzschild case inside the cusp region.

The same observation to the behavior of $\hat{\alpha}$ can be concluded for the case for the black hole in the M87 galaxy, except that it provides a greater value for the weak deflection angle as compared in Sgr. A*. That being said, the M87 galaxy is a good laboratory for weak deflection angle measurements. We emphasize, however, that the solid black curve implies that the receiver must be inside the galaxy itself since it models finite distance. Thus, the one applicable is the solid red curve, representing Eq. \eqref{e20} which is valid for any receiver outside and far from M87. Finally, we note that in the CDM profile, we used the values of the following parameters: $\rho_c=0.008$x$10^{7.5}M_\odot\text{kpc}^{-3}$, $r_\text{c} = 130\text{ kpc}$, and $m = 6.50$x$10^{9}M_\odot$ \cite{Jusufi_2019} for Fig. \ref{fig1n}.

\subsection{Effect of the SFDM profile on weak deflection angle by black holes} \label{sec3}
The CDM profile is known to exhibit a cusp phenomenon near the black hole, around $1$ kpc and below in particular. The Scalar Field Dark Matter model resolves this and one of the known profiles incorporates the Bose-Einstein Condensate (BEC) \cite{Boehmer:2007um}. In such a profile, the density is given as
\begin{equation} \label{e21}
 \rho=\frac{\rho_c r_\text{c}}{\pi r}\sin\left(\frac{\pi r}{r_\text{c}}\right).
\end{equation}
The metric function $A(r)$ of a black hole in SFDM profile then takes the form
\begin{equation} \label{e22}
    A(r)=\exp\left[-\frac{8k\sin\left(\frac{\pi r}{r_\text{c}}\right)}{\pi^2 r}\right]-\frac{2m}{r}.
\end{equation}
Using Eq. \eqref{e4}, we find
\begin{equation} \label{e23}
    F(u) = \frac{1}{b^{2}} - u^{2} + 2mu^{3} + \frac{8 u^3 k}{\pi^2} \sin\left(\frac{\pi}{r_\text{c} u}\right).
\end{equation}
It turns out that the implementation of the perturbation method to Eq. \eqref{e23} still gives the approximate solution similar to Eq. \eqref{e19n}. After solving Eq. \eqref{e23} iteratively, we obtain the closest approach as
\begin{align} \label{e24}
    u_o&=\frac{\sin\phi}{b}+\frac{m\left(1+\cos^{2}\phi\right)}{b^{2}} + \frac{4k}{b^2\pi^2}\sin\left(\frac{b\pi}{r_\text{c}}\right)
\end{align}
Using this, we find $\phi_{\text{RS}}$ as
\begin{widetext}
\begin{align} \label{e25}
    \phi_{\text{RS}}&=(\phi_{\text{RS}})_{\text{Schw}}+\frac{4k\sin\left(\frac{\pi b}{r_\text{c}}\right)}{\pi^{2}b}\left[\frac{1}{\sqrt{1-b^{2}u_{\text{R}}^{2}}}+\frac{1}{\sqrt{1-b^{2}u_{\text{S}}^{2}}}\right] \nonumber\\
    &+\frac{4bkm}{\pi^{2}}\sin\left(\frac{\pi b}{r_\text{c}}\right)\left[\frac{u_{\text{R}}^{3}}{\left(1-b^{2}u_{\text{R}}^{2}\right)^{3/2}}+\frac{u_{\text{S}}^{3}}{\left(1-b^{2}u_{\text{S}}^{2}\right)^{3/2}}\right]+\mathcal{O}(m^2,k^2,m^2k^2),
\end{align}
and for the last two terms in Eq. \eqref{e3}, we find
\begin{align} \label{e26}
    \Psi_{\text{R}}-\Psi_{\text{S}}&=(\Psi_{\text{R}}-\Psi_{\text{S}})_{\text{Schw}}-\frac{4bk}{\pi^{2}}\left[\frac{u_{\text{R}}^{2}\sin\left(\frac{\pi}{r_\text{c} u_{\text{R}}}\right)}{\sqrt{1-b^{2}u_{\text{R}}^{2}}}+\frac{u_{\text{S}}^{2}\sin\left(\frac{\pi}{r_\text{c} u_{\text{S}}}\right)}{\sqrt{1-b^{2}u_{\text{S}}^{2}}}\right] \nonumber\\
    &+\frac{4bkm}{\pi^{2}}\left[\frac{u_{\text{R}}^{3}\left(2b^{2}u_{\text{R}}^{2}-1\right)\sin\left(\frac{\pi}{r_\text{c} u_{\text{R}}}\right)}{\left(1-b^{2}u_{\text{R}}^{2}\right)^{3/2}}+\frac{u_{\text{S}}^{3}\left(2b^{2}u_{\text{S}}^{2}-1\right)\sin\left(\frac{\pi}{r_\text{c} u_{\text{S}}}\right)}{\left(1-b^{2}u_{\text{S}}^{2}\right)^{3/2}}\right]+\mathcal{O}(m^2,k^2,m^2k^2).
\end{align}
Combining the two previous equations above, the weak deflection angle with $u$ being finite leads to
\begin{align} \label{e27}
    \hat{\alpha}&=\frac{2m}{b}\left(\sqrt{1-b^{2}u_{\text{R}}^{2}}+\sqrt{1-b^{2}u_{\text{S}}^{2}}\right)-\frac{4k}{\pi^{2}b}\Bigg\{ \frac{\left[b^{2}u_{\text{R}}^{2}\sin\left(\frac{\pi}{r_\text{c} u_{\text{R}}}\right)-\sin\left(\frac{\pi b}{r_\text{c}}\right)\right]}{\sqrt{1-b^{2}u_{\text{R}}^{2}}}+\frac{\left[b^{2}u_{\text{S}}^{2}\sin\left(\frac{\pi}{r_\text{c} u_{\text{S}}}\right)-\sin\left(\frac{\pi b}{r_\text{c}}\right)\right]}{\sqrt{1-b^{2}u_{\text{S}}^{2}}}\Bigg\} \nonumber\\
    &+\frac{4bkm}{\pi^{2}}\Bigg\{ \frac{u_{\text{R}}^{3}\left[(2b^{2}u_{\text{R}}^{2}-1)\sin\left(\frac{\pi}{r_\text{c} u_{\text{R}}}\right)+\sin\left(\frac{\pi b}{r_\text{c}}\right)\right]}{\left(1-b^{2}u_{\text{R}}^{2}\right)^{3/2}}+\frac{u_{\text{S}}^{3}\left[(2b^{2}u_{\text{S}}^{2}-1)\sin\left(\frac{\pi}{r_\text{c} u_{\text{S}}}\right)+\sin\left(\frac{\pi b}{r_\text{c}}\right)\right]}{\left(1-b^{2}u_{\text{S}}^{2}\right)^{3/2}}\Bigg\}+\mathcal{O}(m^2,k^2,m^2k^2).
\end{align}
\end{widetext}
Again, if $k=0$ and in the far approximation, $\hat{\alpha}=4m/b$ is recovered. We notice also that $u$ cannot be equal to zero for $\sin(\pi/r_\text{c} u)$. However, with $u$ having a very small value, $\sin(\pi/r_\text{c} u)$ can only be somewhere between $-1$ and $1$. Moreover, since there is a factor of $u^2$ in $\sin(\pi/r_\text{c} u)$, we can safely approximate the term to zero. Therefore, Eq. \eqref{e27} can be reduced to
\begin{equation} \label{e28}
    \hat{\alpha}=\frac{4m}{b}+\frac{8 k}{\pi^2 b}\sin\left(\frac{\pi b}{r_\text{c}}\right).
\end{equation}

\begin{figure} [htpb]
    \centering
    \includegraphics[width=\linewidth]{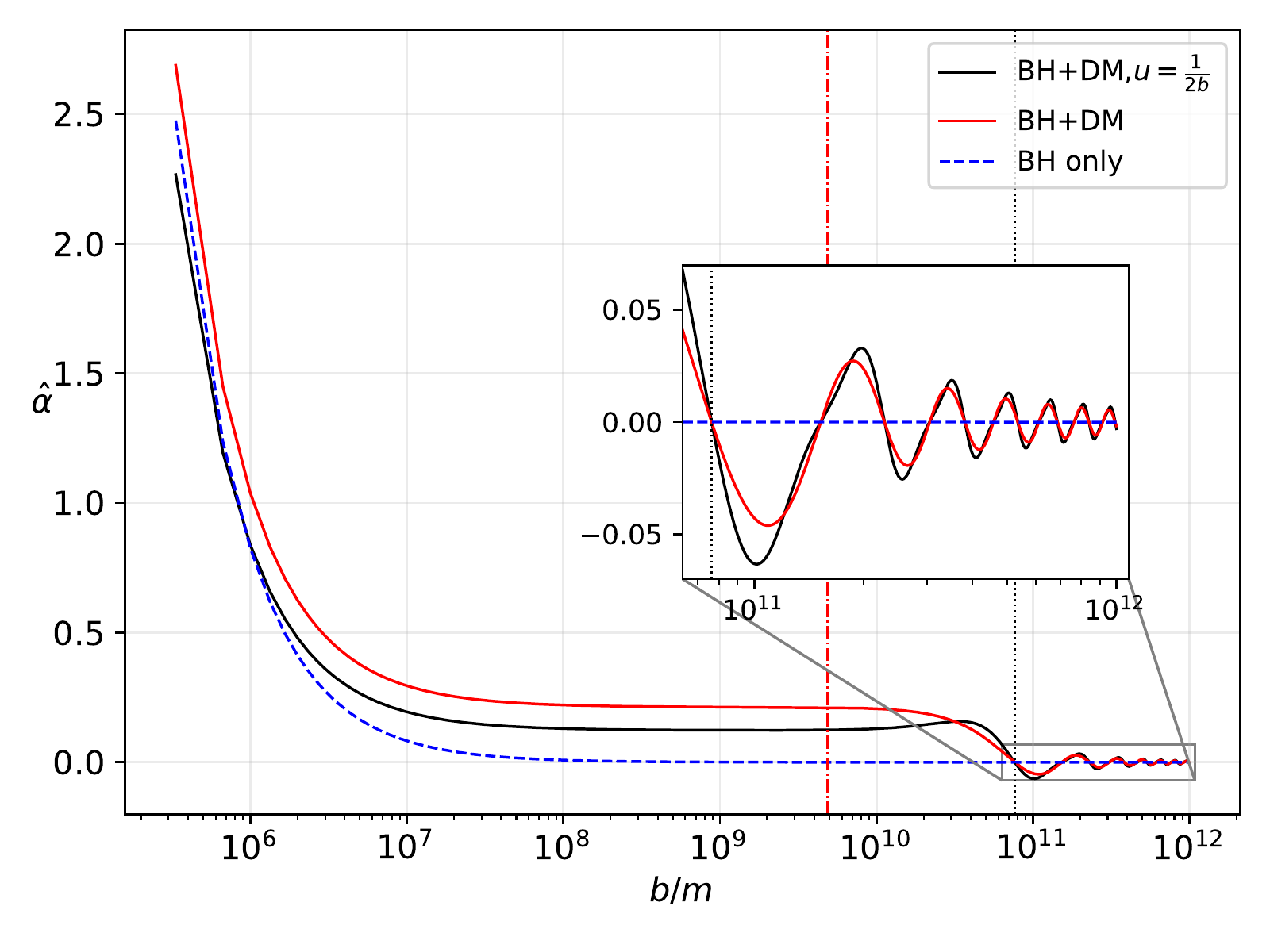}
    \caption{Curve of the weak deflection angle in the SFDM profile for different values of the dimensionless impact parameter $b/m$ in Sgr. A*. The vertical dotted line is the value of the outermost core radius $r_\text{c}\sim7.639$x$10^{10}m$. The region below the dash-dotted vertical line is the cusp region ($\sim 4.86$x$10^9m$).}
    \label{fig2}
\end{figure}
In the SFDM profile, the parameters for Sgr. A* are the following: $\rho_c=3.43$x$10^7M_\odot\text{kpc}^{-3}$ and $r_\text{c} = 15.7\text{ kpc}$ for the core density and radius respectively, and $k = 30869m$. Fig. \ref{fig2} shows how the weak deflection angle varies for different values of $b/m$. For the black hole only case, we can see that the increase in $\hat\alpha$ is very small overall, even as $b/m$ decreases further. The same observation holds when the effect of the SFDM profile is present. Interesting behavior for the $\hat\alpha$ curve occurs for $b>r_\text{c}$ where fluctuation between attractive and repulsive behavior occurs. Furthermore, the amplitude of the fluctuation decreases as $b/m$ increases, which indicates that it reaches the Schwarzschild case for very large $b/m$. The periodic behavior diminishes inside the core radius while we see a considerable deviation. When $b$ is considerable small in comparison to $r_\text{c}$, then $\sin(\pi b/r_\text{c})\sim \pi b/r_\text{c}$. Then the second term becomes $8k/\pi r_\text{c}$ which is a constant. With the parameters used herein, this constant is $\sim1.029$x$10^{-6}$. For this reason, the first term dominates which explains how the dark matter contribution follows the same trend of the Schwarzschild case. In such a region, the only effect of the dark matter in the SFDM profile is to increase slightly the value of the weak deflection angle by some constant that depends on the black hole and dark matter parameters.

Interestingly, we can also see the difference in the behavior of the curves between the finite distance case and the asymptotic one. Outside the core radius, there is a very small difference between the two although the weak deflection angle is quite larger in the finite distance case. Detection of this oscillatory behavior near the core radius can potentially rule out the CDM or the URC profiles. Now, inside the core radius, it can be seen that the asymptotic case for the receiver is better since $\hat{\alpha}$ is larger. As can be gleaned from Fig. \ref{fig2}, the Schwarzschild case for the deflection angle is even better compared to the finite distance case when $b/m$ gets smaller in comparison to the core radius.

\subsection{Effect of the URC profile on weak deflection angle by black holes} \label{sec5}
The Universal Rotation Curve profile came from the another known model called Burkert profile \cite{Castignani:2012sr,Salucci:2007tm}. The URC density profile is expressed as
\begin{equation} \label{e29}
    \rho=\frac{\rho_c r_\text{c}^3}{(r+r_\text{c})(r^2+r_\text{c}^2)}.
\end{equation}
Taking consideration of the black hole, it was shown in \cite{Jusufi_2019} that $A(r)$ is somewhat complicated:
\begin{align} \label{e30}
    A(r)&=\left(1+\frac{r^{2}}{r_\text{c}^{2}}\right)^{-\frac{2\pi k\left(1-\frac{r}{r_\text{c}}\right)}{r}}\left(1+\frac{r}{r_\text{c}}\right)^{-\frac{4\pi k\left(1+\frac{r}{r_\text{c}}\right)}{r}} \times \nonumber\\
    &\exp\left[\frac{4\pi k\left(1+\frac{r}{r_\text{c}}\right)\arctan\left(\frac{r}{r_\text{c}}\right)}{r}\right]-\frac{2m}{r}.
\end{align}
Nevertheless, it can be easily shown that when $k=0$, the first term above is equal to $1$.
Calculating the orbit equation we find
\begin{align} \label{e31}
    &F(u)= \frac{1}{b^{2}} - u^{2} + 2mu^{3} + 8 \pi u^3 k \ln\left(1+\frac{1}{ur_\text{c}}\right) \nonumber\\
    &-\frac{2\pi ku^{2}}{r_{\text{c}}}\Bigg\{ (r_{\text{c}} u+1)\left[2\arctan\left(\frac{1}{r_{\text{c}} u}\right)-\ln\left(1+\frac{1}{r_{\text{c}}^{2}u^{2}}\right)\right] \nonumber\\
    &-2\ln\left(1+\frac{1}{r_{\text{c}} u}\right)\Bigg\},
\end{align}
and notice how we recovered the CDM contribution in the fourth term of the equation above despite not having Eq. \eqref{e12} in Eq. \eqref{e30}. One can verify that the approximate solution again gives the expression in Eq. \eqref{e19n}. After we solve Eq. \eqref{e31} iteratively, the inverse of the closest approach is found as
\begin{equation} \label{e32}
    u_o=\frac{\sin\phi}{b}+\frac{m\left(1+\cos^{2}\phi\right)}{b^{2}}+\frac{4\pi k}{b^{2}}\ln\left(1+\frac{b}{r_\text{c}}\right)+\frac{\pi k}{r_{\text{c}} b^2}\lambda,
\end{equation}
where
\begin{align} \label{e39n}
    \lambda&=(r_{\text{c}}-b)\ln\left(b^{2}+r_{\text{c}}^{2}\right)+2b\ln(b+r_{\text{c}}) \nonumber\\
    &-2(b+r_{\text{c}})\arctan\left(\frac{b}{r_{\text{c}}}\right)-2r_{\text{c}}\ln(r_{\text{c}}).
\end{align}
Using Eq. \eqref{e32}, we solve $\phi$ and obtain $\phi_{\text{RS}}$ as
\begin{widetext}
\begin{align} \label{e33}
    \phi_{\text{RS}}&=(\phi_{\text{RS}})_{\text{Schw+CDM}}+\frac{\pi k\lambda}{br_{\text{c}}}\left(\frac{1}{\sqrt{1-b^{2}u_{\text{R}}^{2}}}+\frac{1}{\sqrt{1-b^{2}u_{\text{S}}^{2}}}\right) \nonumber\\
    &+\frac{2\pi k\lambda m}{br_{\text{c}}}\left(\frac{u_{\text{R}}}{\left(1-b^{2}u_{\text{R}}^{2}\right)^{3/2}}+\frac{u_{\text{S}}}{\left(1-b^{2}u_{\text{S}}^{2}\right)^{3/2}}\right)+\mathcal{O}(m^2,k^2,m^2k^2).
\end{align}
For $\Psi_R-\Psi_S$, we have
\begin{align} \label{e35}
    \Psi_{\text{R}}-\Psi_{\text{S}}&=(\Psi_{\text{R}}-\Psi_{\text{S}})_{\text{Schw+CDM}}+\frac{2\pi bk}{r_{\text{c}}}\left(\frac{\gamma_{\text{R}} u_{\text{R}}}{\sqrt{1-b^{2}u_{\text{R}}^{2}}}+\frac{\gamma_{\text{S}} u_{\text{S}}}{\sqrt{1-b^{2}u_{\text{S}}^{2}}}\right) \nonumber\\
    &-\frac{2\pi bkm}{r_{\text{c}}}\left(\frac{\gamma_{\text{R}} u_{\text{R}}^{2}\left(2b^{2}u_{\text{R}}^{2}-1\right)}{\left(1-b^{2}u_{\text{R}}^{2}\right)^{3/2}}+\frac{\gamma_{\text{S}} u_{\text{S}}^{2}\left(2b^{2}u_{\text{S}}^{2}-1\right)}{\left(1-b^{2}u_{\text{S}}^{2}\right)^{3/2}}\right)+\mathcal{O}(m^2,k^2,m^2k^2),
\end{align}
where we have written
\begin{equation}
	\gamma=\frac{1}{2}\left[(1-r_{\text{c}} u)\ln\left(1+\frac{1}{r_{\text{c}}^{2}u^{2}}\right)-2\ln\left(1+\frac{1}{r_{\text{c}} u}\right)+2\arctan\left(\frac{1}{r_{\text{c}} u}\right)\right].
\end{equation}
Combining the above equations, we find weak deflection angle in finite distance as
\begin{align} \label{e37}
    \hat{\alpha}&=\hat{\alpha}_{\text{Schw+CDM}}+\frac{\pi k}{br_{\text{c}}}\left[\frac{\left(2b^{2}\gamma_{\text{R}} u_{\text{R}}+\lambda\right)}{\sqrt{1-b^{2}u_{\text{R}}^{2}}}+\frac{\left(2b^{2}\gamma_{\text{S}} u_{\text{S}}+\lambda\right)}{\sqrt{1-b^{2}u_{\text{S}}^{2}}}\right] \nonumber\\
&-\frac{2\pi km}{br_{\text{c}}}\left[\frac{u_{\text{R}}\left(2b^{4}\gamma_{\text{R}} u_{\text{R}}^{3}-b^{2}\gamma_{\text{R}} u_{\text{R}}-\lambda\right)}{\left(1-b^{2}u_{\text{R}}^{2}\right)^{3/2}}+\frac{u_{\text{S}}\left(2b^{4}\gamma_{\text{S}} u_{\text{S}}^{3}-b^{2}\gamma_{\text{S}} u_{\text{S}}-\lambda\right)}{\left(1-b^{2}u_{\text{S}}^{2}\right)^{3/2}}\right]+\mathcal{O}(m^2,k^2,m^2k^2),
\end{align}
\end{widetext}
where the first term is expressed by Eq. \eqref{e19}. As expected, the Schwarzschild case is recovered if $k=0$. One can not set $u=0$ due to the divergences in the $\ln$ and $\arctan$ terms. We say that these are only apparent divergences. Nonetheless, if $u$ is very small, these terms can have finite values. If this is the case, then $u$ and $u^2$ as scaling factors will dominate and we can safely assume that in the far approximation, Eq. \eqref{e37} should reduce to
\begin{equation} \label{e38}
    \hat{\alpha}=\frac{4m}{b}+\frac{8\pi k}{b}\ln\left(1+\frac{b}{r_\text{c}}\right)+\frac{2\pi k \lambda}{b r_{\text{c}}}
\end{equation}
It is interesting how we can see the emergence of the CDM profile contribution in Eq. \eqref{e38}.
\begin{figure} [htpb]
    \centering
    \includegraphics[width=\linewidth]{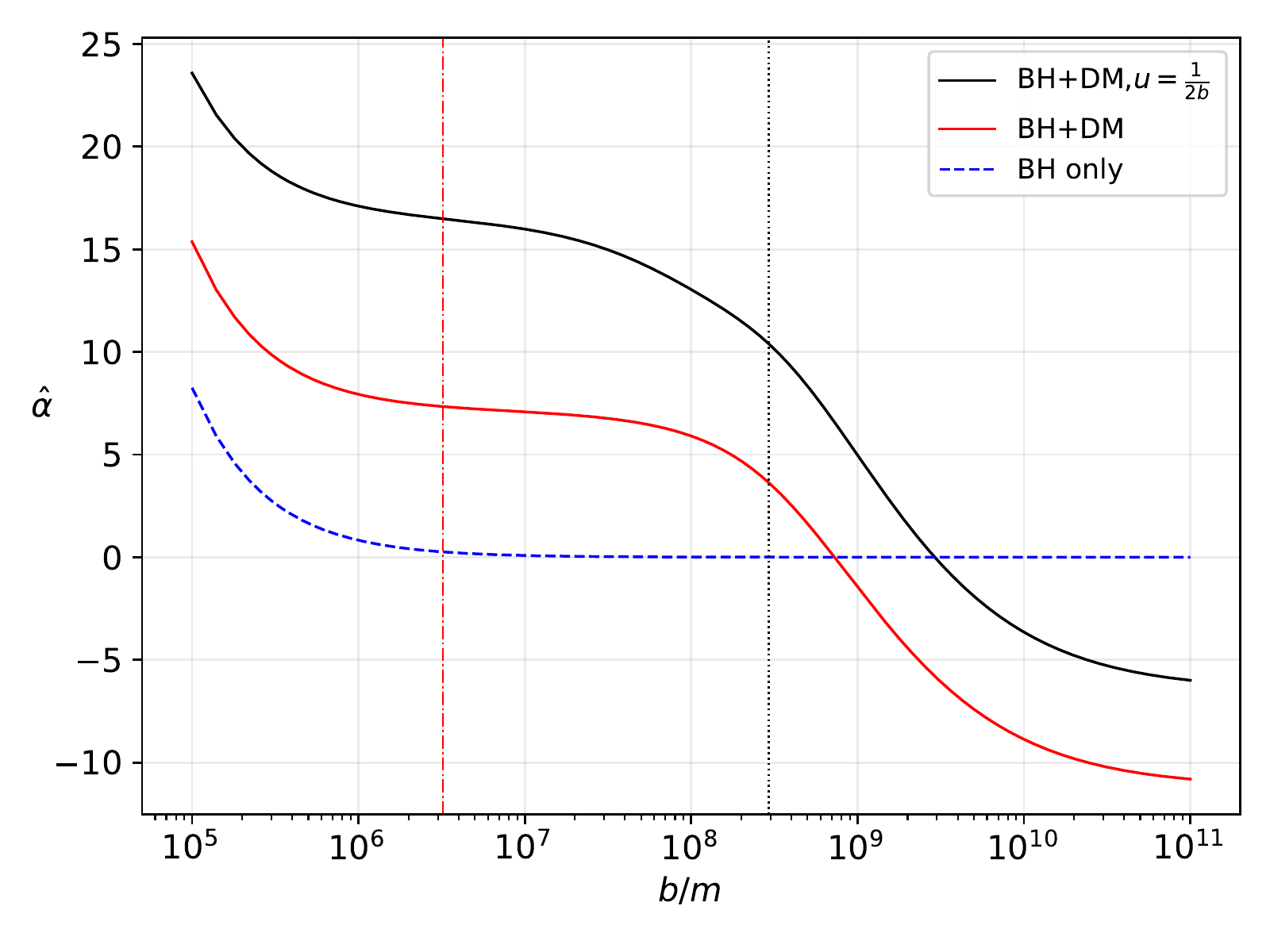}
    \caption{Curve of the weak deflection angle in the URC profile for different values of impact parameter $b/m$ in M87. The vertical dotted line is the value of the outermost core radius $r_\text{c}\sim2.932$x$10^{8}m$. The region below the dash-dotted vertical line is the cusp region ($\sim 3.214$x$10^6m$).}
    \label{fig3}
\end{figure}
\begin{figure} [htpb]
    \centering
    \includegraphics[width=\linewidth]{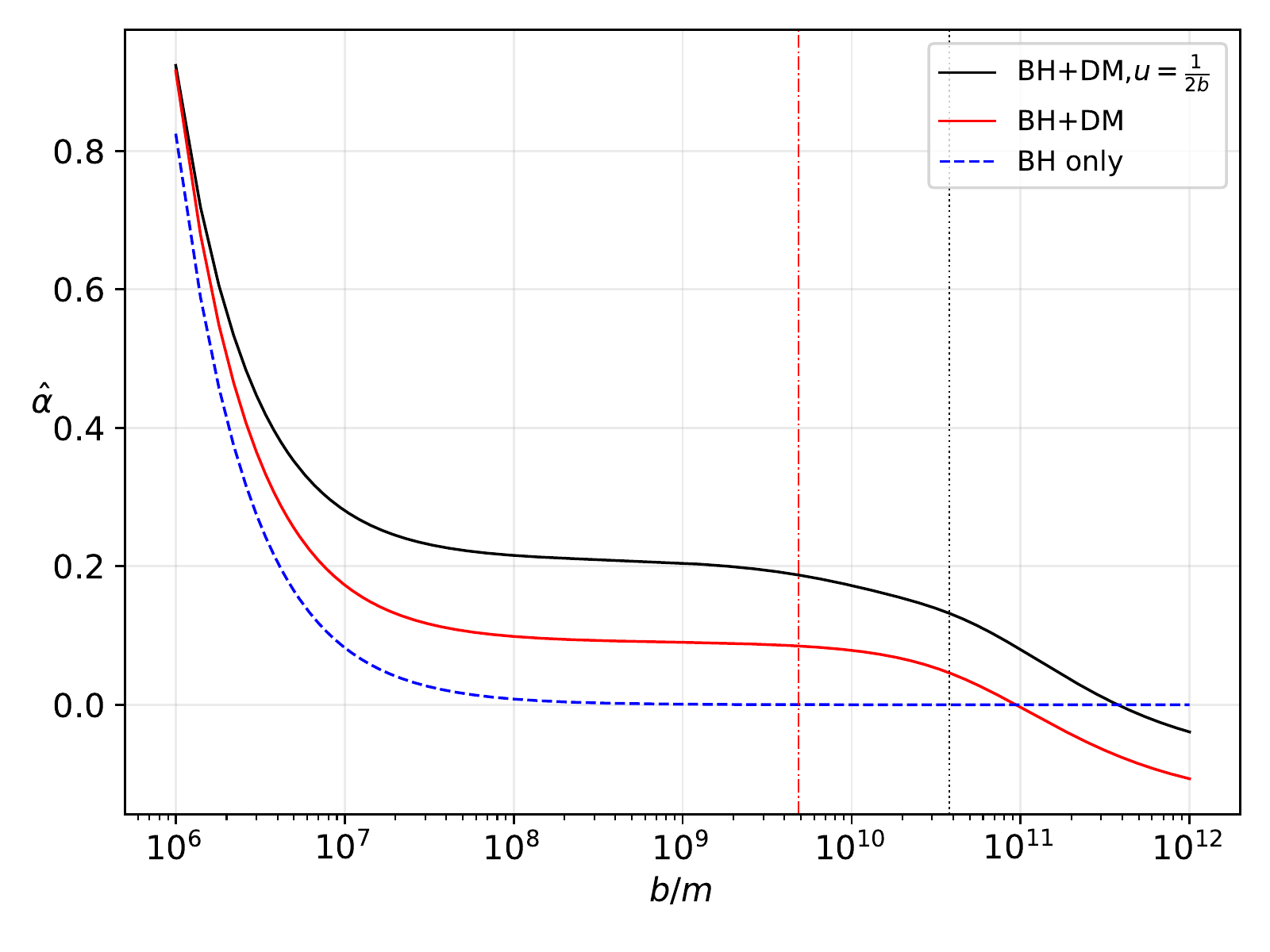}
    \caption{Curve of the weak deflection angle in the URC profile for different values of impact parameter $b/m$ in Sgr. A*. The vertical dotted line is the value of the outermost core radius $r_\text{c}\sim3.79$x$10^{10}m$. The region below the dash-dotted vertical line is the cusp region ($\sim 4.859$x$10^9m$).}
    \label{fig3n}
\end{figure}
Let us use the M87 as an example \cite{Jusufi_2019} to demonstrate the effect of the URC profile on the weak deflection angle. The mass of the black hole at the center is given as $m=6.5$x$10^9M_\odot$, while the dark matter parameters are $\rho_c=6.9$x$10^6M_\odot\text{kpc}^{-3}$ and $r_\text{c} = 91.2\text{ kpc}$ for the core density and radius respectively. We can then use $k = 805m$ for the dark matter mass. In Fig. \ref{fig3}, we can see some interesting behavior of the $\hat{\alpha}$ curve between the three cases. First, we observe that the deflection angle outside the core radius is negative, which implies that photons are deflected repulsively by the amalgamation of the dark matter and black hole. We remark that the repulsive deflection angle is nothing new since in Ref. \cite{Nakashi2019}, the deflection angle due to the black hole in $(2+1)$D massive gravity also sometimes gives a negative value. We examined the extension of this plot for the case when $b>>r_{\text{c}}$, and one can verify that the curves do not approach the Schwarzschild case. However, despite giving a repulsive deflection, this case favors the receivers at $u \rightarrow 0$ than the one in the finite distance. We see a steep increase in $\hat{\alpha}$ as the impact parameter approaches the value of the core radius, and it levels off inside. Beyond the cusp region, we can see again that the dark matter effect follows the same trend as the Schwarzschild case, implying that in such a region, the CDM term dominates. We should also note that exploring the weak deflection angle inside the dark matter core radius would be more detectable to a receiver with a finite distance from the black hole. We also plotted the URC profile for Sgr. A* using $\rho_c=1.2$x$10^7M_\odot\text{kpc}^{-3}$ and $r_\text{c} = 7.8\text{ kpc}$ for the core density and radius respectively \cite{Lin2019}. In these parameters, $k=0.9m$, which we can see how it compares to M87. Thus, a greater dark matter mass for M87 explains why such a galaxy can be useful for dark matter detection using the weak deflection angle. The same conclusion can be seen in Fig. \ref{fig3n}, and we can see clearly how the dark matter effect for $\hat{\alpha}$ behaves similar to the Schwarzschild case deep within the center of the galaxy.

\subsection{Remarks on the cusp phenomenon}
In this section, we comment on the possible effect of the cusp phenomenon on the weak deflection angle. Or, is there any? Looking at Fig. (1) in Ref. \cite{Xu_2018}, we can see that the energy density is still finite at $r=1$ kpc and it is indeed asymptotic to $+\infty$ as $r$ approaches zero. Not taking into account the black hole, only the CDM profile produces a cusp, while the SFDM has none. With the black hole into consideration, it is reported \cite{Xu_2018} that the cusp arises in the SFDM profile as the energy density is enhanced near the black hole. We see that this cusp if it begins at $r=1$ kpc, it does not affect the weak deflection angle as can be gleaned from Figs. \ref{fig1}-\ref{fig2}. The reason is that there is no peculiar deviation to the curve happening at $r=1$ kpc and the increase in $\hat{\alpha}$ happens even without the influence of the dark matter profile. The weak deflection angle with the CDM and SFDM profiles merely follows the same trend as the Schwarzschild case. In Ref. \cite{Jusufi_2019}, it is reported that the black hole with the URC profile has no cusp phenomenon since the energy density is finite at very low values of $r$. We then see in Figs. \ref{fig3} and \ref{fig3n} the behavior of the $\hat{\alpha}$ curve without the effect of the cusp. With these observations from the three profiles, we cannot attribute the behavior of $\hat{\alpha}$ inside the core radius to the cusp phenomenon, but instead to the kind of astrophysical environment where the light travels. The effect of the cusp may be more evident very close to the black hole and for this reason, a study of the strong deflection of light is necessary to investigate such effect. Related to this, it was already reported that the horizon, ergosphere, shadow radius, and energy emission rate exhibit deviations to the standard values, and cusp effects may be one of the reasons for such deviations \cite{Xu_2018,Jusufi_2019,Hou_2018a}. The conclusion is reasonable since these observables happen near the black hole where their fundamental properties are maintained despite being surrounded by dark matter.

\section{Effect of the dark matter profiles to the Einstein Ring}
Let us now calculate and form an estimate as to what will be the angular size of the Einstein rings due to the effect of the dark matter density profiles in this study. First, let $D_\text{S}$ and $D_\text{R}$ be the distance of the source and the receiver respectively from the lensing object. Considering thin lens equation, $D_\text{RS}=D_\text{R}+D_\text{S}$.
To find the position of the weak field images, we consider the lens equation given as \cite{Bozza2008}
\begin{equation}
    D_\text{RS}\tan\beta=\frac{D_\text{R}\sin\theta-D_\text{S}\sin(\hat{\alpha}-\theta)}{\cos(\hat{\alpha}-\theta)}.
\end{equation}
An Einstein ring is formed under the condition $\beta=0$, and the above equation gives the ring's angular radius as
\begin{equation}
    \theta_\text{E}\sim\frac{D_\text{S}}{D_\text{RS}}\hat{\alpha}.
\end{equation}
Furthermore, if the Einstein ring is assumed to be very small, we can use the relation $b=D_\text{R}\sin\theta \sim D_\text{R}\theta$. We also note that the terms in Eq. \eqref{e20} are already in the leading order, which are dependent on the BH and dark matter mass. Let us also denote $b/r_\text{c}=\epsilon$. We then find the angular size of the Einstein ring in the CDM profile as
\begin{equation} \label{e47}
    \theta_\text{E}^{\text{CDM}}=2\sqrt{\frac{D_\text{S}}{D_\text{RS}D_\text{R}}\left[m+2\pi k \ln \left(1+\epsilon\right)\right]}.
\end{equation}
Meanwhile, we found the angular radius for the SFDM and URC as
\begin{equation}
    \theta_\text{E}^{\text{SFDM}}=2\sqrt{\frac{D_\text{S}}{D_\text{RS}D_\text{R}}\left[m+\frac{2k}{\pi^2}\sin(\pi \epsilon)\right]},
\end{equation}
\begin{equation} \label{e49}
    \theta_\text{E}^{\text{URC}}=2\sqrt{\frac{D_\text{S}}{D_\text{RS}D_\text{R}}\left[m+2\pi k \ln \left(1+\epsilon\right)+\frac{\pi k \lambda}{2 r_\text{c}}\right]}
\end{equation}
respectively, where $\lambda$ is given by Eq. \eqref{e39n}. From the galactic center to our location, the distance is approximately $D_\text{R}\sim 8.3$ kpc, which justifies the use of the equations \eqref{e20}, \eqref{e28}, and \eqref{e38}. Without the influence of dark matter, the angular radius depends solely to the source's distance from the lensing object if $D_\text{R}$ is fixed.

\begin{figure} [htpb]
    \centering
    \includegraphics[width=\linewidth]{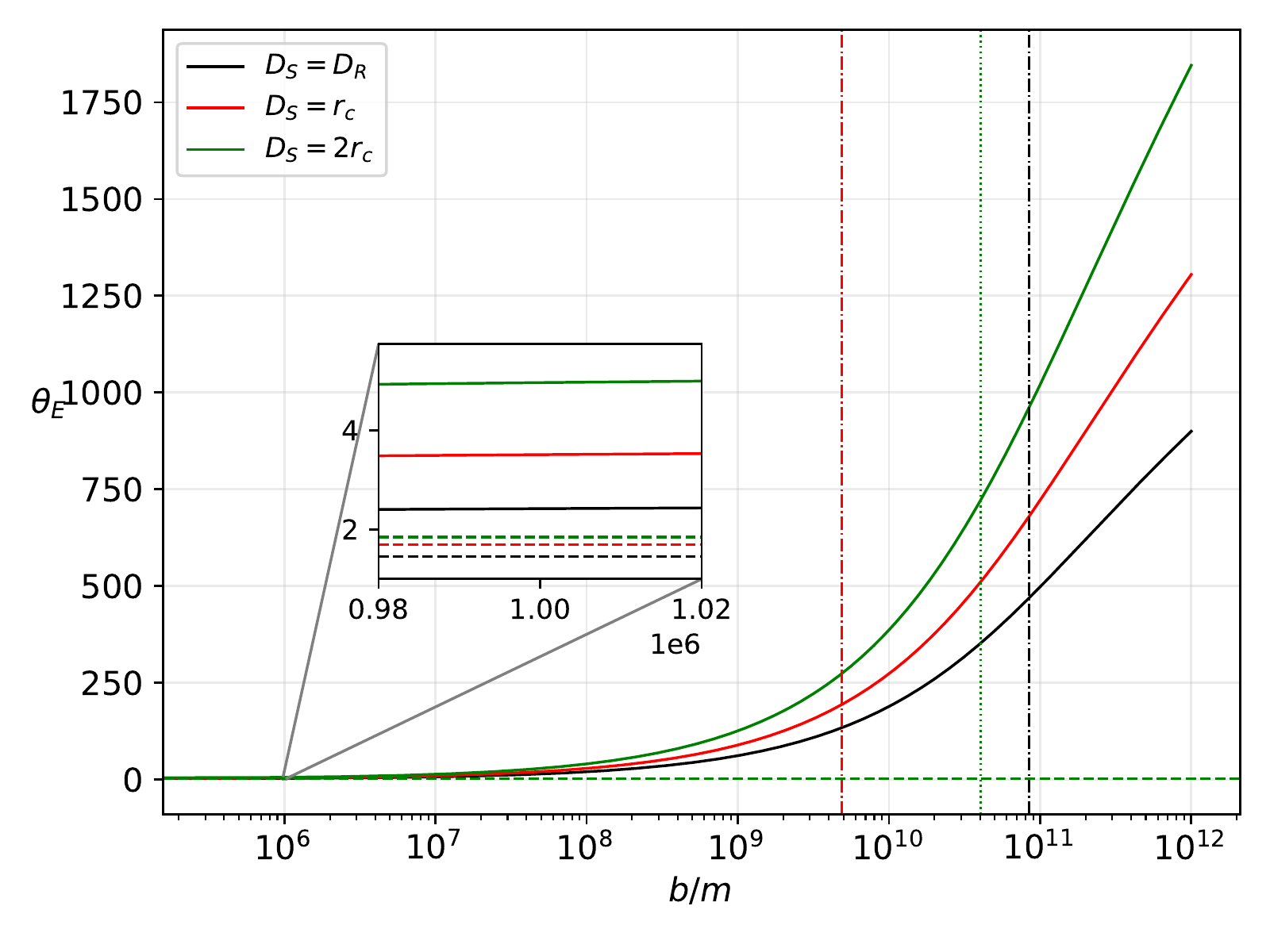}
    \caption{The angular radius of the Einstein ring for CDM profile in Sgr. A*. Here, $\theta_\text{E}$ is in $\mu$as. We used the same parameters present in the weak deflection angle plots, except for $D_\text{R} = 8.3$ kpc, represented by the green dotted line.}
    \label{fig7}
\end{figure}

\begin{figure} [htpb]
    \centering
    \includegraphics[width=\linewidth]{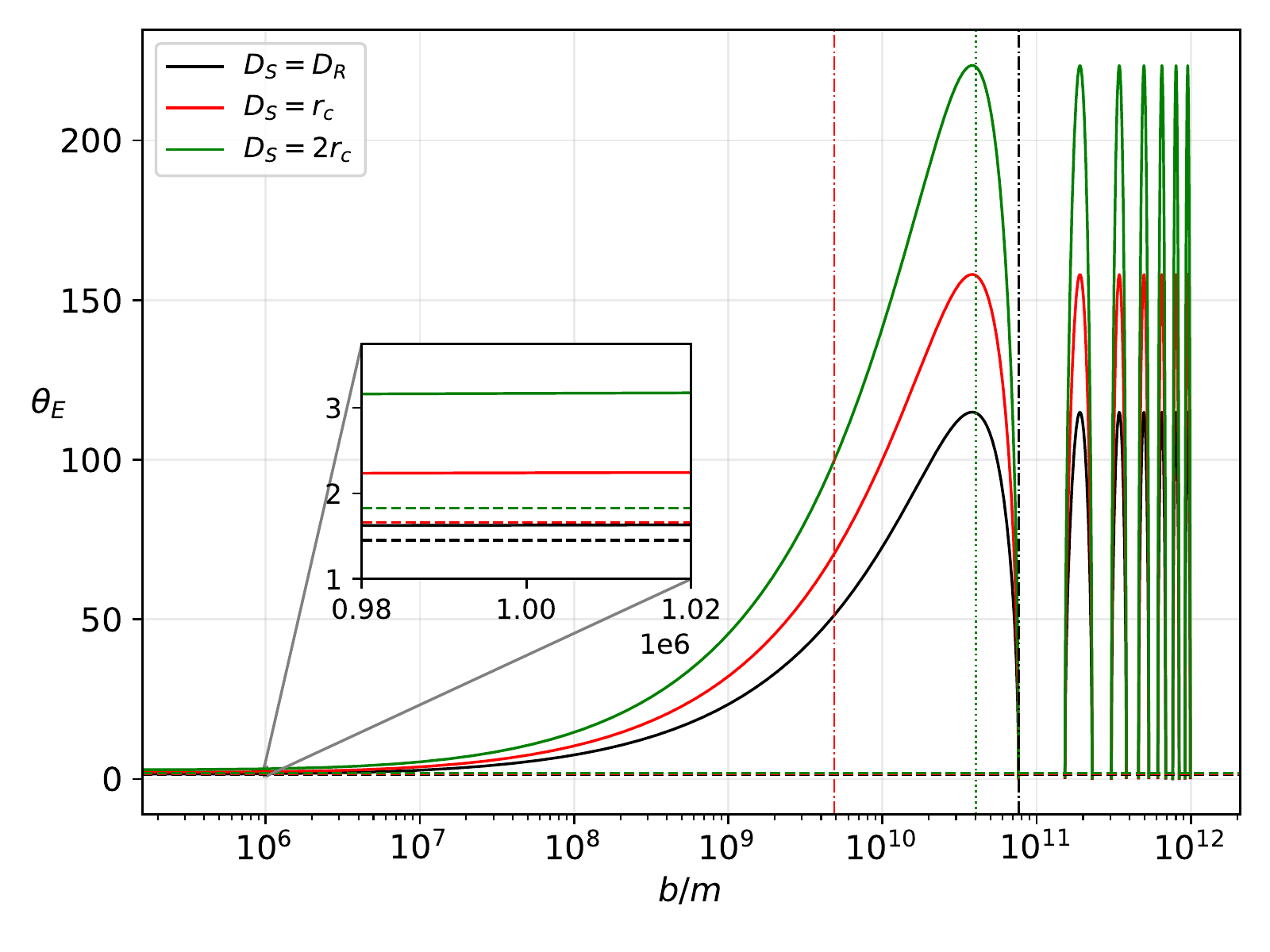}
    \caption{The angular radius of the Einstein ring for SFDM profile in Sgr. A*}
    \label{fig8}
\end{figure}

\begin{figure} [htpb]
    \centering
    \includegraphics[width=\linewidth]{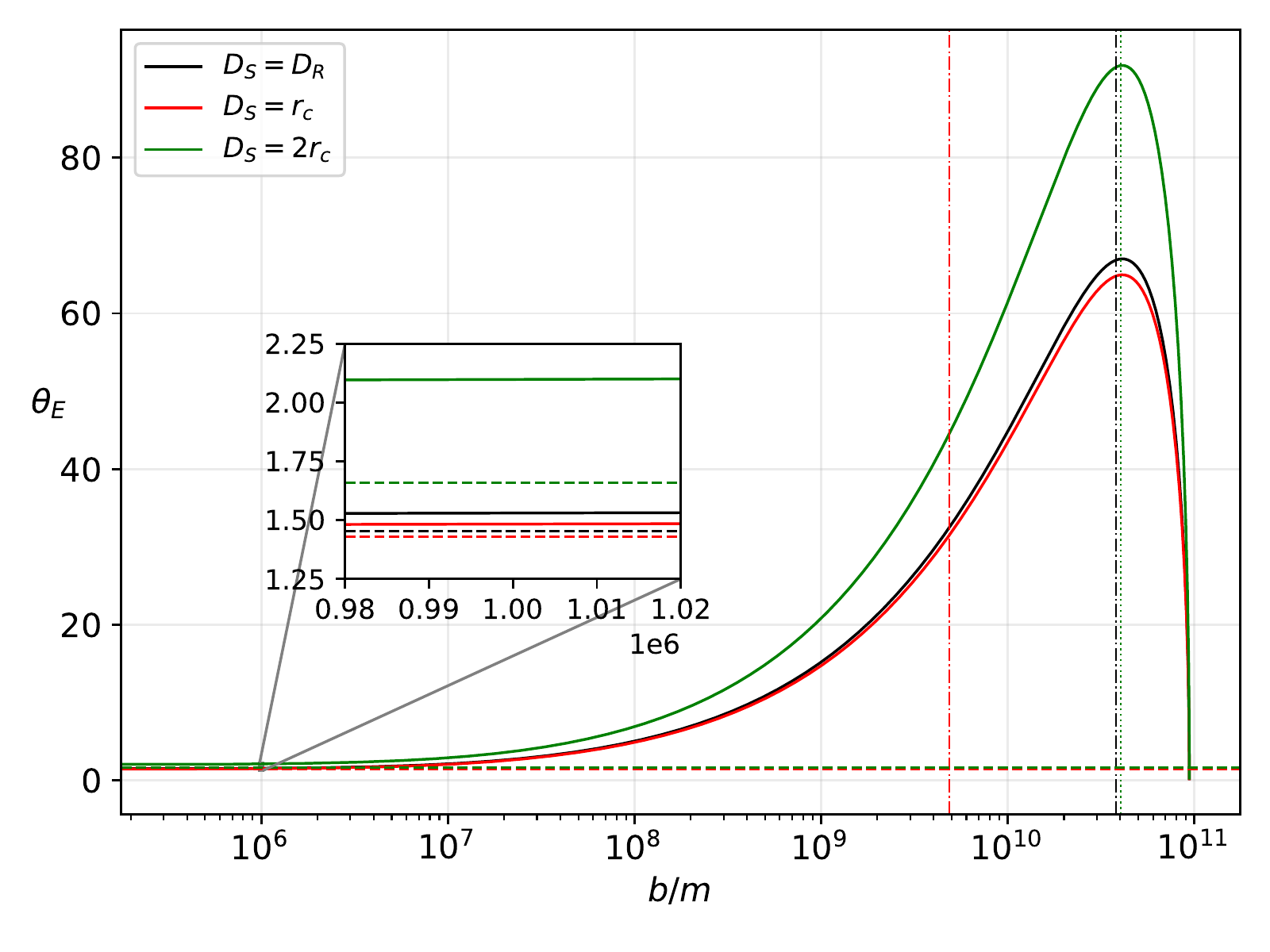}
    \caption{The angular radius of the Einstein ring for URC profile in Sgr. A*}
    \label{fig9}
\end{figure}

It is known that dark matter effects are more obvious far from the galactic center. For this reason, we plotted Eqs. \eqref{e47}-\eqref{e49} showing how the distance of the source and the impact parameter will affect the size of the Einstein ring. To be more realistic, we have used the Sgr. A* and the estimated distance of the solar system from the galactic center. See Figs. \ref{fig7}-\ref{fig9}. For comparison, we have also shown the Schwarzschild case represented by the dashed line for each source location. As we have seen in the inset plot, near the galactic center, the deviation is small but tends to increase the Einstein ring angular radius. It is a finding that is in contrast with the results in Ref. \cite{Atamurutov2022} using the perfect fluid dark matter model, where the radius of the Einstein ring decreases as the impact parameter of light increases. However, the effect of varying the distance of the source from the lensing object was not explored.

An interesting behavior of the curve happens near the core radius. For the CDM profile, we can see a continuous increase in the radius, and the steepness of the curve decreases for $b/m>10^{12}$ (not shown). For the SFDM profile, the peak of the curve arises near $b/m = D_\text{R}$, then a steep decrease in radius is seen. We see that the deviation went back to the Schwarzschild case near the core radius. For any $b/m>r_\text{c}$, we see a fluctuating deviation for the radius. In the URC profile, the peak of $\theta_\text{E}$ is near the core radius, and a fast decline in the angular radius is observed just outside the core radius. With the parameters given for Sgr. A*, there are no Einstein rings observed near and beyond $b/m = 10^{11}$. A possible explanation for this behavior is that beyond this value of $b/m$, the weak deflection angle is negative, which causes a repulsive deflection, not reaching the intended location of the receiver (See Fig. \ref{fig3n}). As a final remark, we have seen that each profile gives a unique behavior for the Einstein ring deviation.

\section{Conclusion} \label{conc}
In this work, we have obtained the weak deflection angle with finite distance for black holes surrounded by dark matter. In particular, we considered the dark matter profiles such as the CDM, SFDM, and URC profiles. With this aim, we calculated the positional angles of the source and the receiver using the method by Ishihara et. al \cite{Ishihara_2016}, and the calculation of their longitudinal angle difference using the non-asymptotic GBT method in Ref. \cite{Li:2020wvn}. Using the known data and parameters for the black hole and dark matter in the Milky Way and M87 galaxies, the density profiles give a unique behavior for the weak deflection angle. We see in Figs. \ref{fig1}-\ref{fig3n} how they differ. Outside the dark matter core radius, the CDM profile exhibits a positive deflection angle and the finite distance gave the highest value for $\hat{\alpha}$. In the URC profile, a negative deflection angle manifests, and the largest value for $\hat{\alpha}$ can be measured by a receiver at $u\rightarrow 0$. We also note that the CDM profile naturally emerges in the URC profile, and its manifestation is evident inside the core radius since these two profiles follow the same behavior. Furthermore, the largest value for $\hat{\alpha}$ is seen by the receiver in the finite distance for impact parameters that are less than the core radius. On the other hand, the SFDM profile gives an interesting behavior to the weak deflection angle since the deviation is oscillatory near the outside core radius, but follows the Schwarzschild trend inside the core radius. Inside the core radius, we can say that $\hat{\alpha}$ behaves differently as compared to the CDM and URC profile because, at this domain, the receiver at $u\rightarrow 0$ reads the highest value for $\hat{\alpha}$. See Fig. \ref{fig2}. Finally, with the three profiles presented in this study, we find that the phenomenon of weak deflection angle tends to ignore the effects of cusp reported to be present in CDM and SFDM. However, one could distinguish the dark matter profile when we observe the weak deflection angle near the core radius. With the differing behavior of the $\hat{\alpha}$ curve, it can give us hints on how dark matter and photons interact, at the domain near the core radius. We believe that such an analysis is better than analyzing the depths of the black hole shadow. Looking again at Figs. \ref{fig1} - \ref{fig3n}, we can see how the dark matter mass is the dominating term for the value of the weak deflection angle. We observe small values for $\hat{\alpha}$ in the CDM profile for M87 as compared to its value in the URC profile. The lowest value is in Sgr. A* in the URC profile.

We have also analyzed how the angular radius of the Einstein ring behaves in each of the profiles. Near the galactic center, the deviation to the Schwarzschild case is small, where the radius is increased. As we also increase the distance of the source and the impact parameter of light to the lensing object, we observe more deviation to the Einstein ring radius. Dark matter effects are also seen to amplify its effect near the core radius. See Figs. \ref{fig7}-\ref{fig9}.

Let us list some space technology that can potentially detect the deviation in the weak deflection angle or the Einstein ring caused by the dark matter density profiles. First, we have the Event Horizon Telescope that can achieve an angular resolution of $10-15 \mu$as within $345$ GHz in the future. In addition, while the ESA GAIA mission is also capable of providing around $20\mu$as - $7\mu$as of angular resolution \cite{Liu_2017}, a more powerful space-based technology called the VLBI RadioAstron \cite{Kardashev2013} can obtain an angular resolution ranging from $1-10\mu$as. We then conclude that they are within the capability of extracting information from the dark matter profiles herein, especially for high dark matter density. Although the weak deflection angle is small for Sgr. A*, the angular radius for the Einstein ring is detectable.

As a future research direction, it is interesting to include black hole rotation in the analysis, which is a work in progress. Another is the comparison between deflection angles of null and time-like particles. The cusp phenomenon is also interesting and the calculation of the strong deflection angle by a black hole in a dark matter profile may reveal its effect. Further investigation is also needed whether the cusp phenomenon affects the black hole geometry, and this can be achieved by comparing a cusp profile to a cored profile.

\bibliography{references}

\begin{thebibliography}{99}
\expandafter\ifx\csname natexlab\endcsname\relax\def\natexlab#1{#1}\fi
\expandafter\ifx\csname bibnamefont\endcsname\relax
  \def\bibnamefont#1{#1}\fi
\expandafter\ifx\csname bibfnamefont\endcsname\relax
  \def\bibfnamefont#1{#1}\fi
\expandafter\ifx\csname citenamefont\endcsname\relax
  \def\citenamefont#1{#1}\fi
\expandafter\ifx\csname url\endcsname\relax
  \def\url#1{\texttt{#1}}\fi
\expandafter\ifx\csname urlprefix\endcsname\relax\def\urlprefix{URL }\fi
\providecommand{\bibinfo}[2]{#2}
\providecommand{\eprint}[2][]{\url{#2}}

\bibitem[{\citenamefont{Schwarzschild}(1916)}]{Schwarzschild_1916}
\bibinfo{author}{\bibfnamefont{K.}~\bibnamefont{Schwarzschild}},
  \bibinfo{journal}{Sitzungsber. Preuss. Akad. Wiss. Berlin (Math. Phys. )}
  \textbf{\bibinfo{volume}{1916}}, \bibinfo{pages}{189} (\bibinfo{year}{1916}).

\bibitem[{\citenamefont{{The Event Horizon Telescope Collaboration, et
  al.}}(2021{\natexlab{a}})}]{Event2021a}
\bibinfo{author}{\bibnamefont{{The Event Horizon Telescope Collaboration, et
  al.}}}, \bibinfo{journal}{Astrophys. J. Lett.}
  \textbf{\bibinfo{volume}{910}}, \bibinfo{pages}{L12}
  (\bibinfo{year}{2021}{\natexlab{a}}).

\bibitem[{\citenamefont{{The Event Horizon Telescope Collaboration, et
  al.}}(2021{\natexlab{b}})}]{Event2021b}
\bibinfo{author}{\bibnamefont{{The Event Horizon Telescope Collaboration, et
  al.}}}, \bibinfo{journal}{Astrophys. J. Lett.}
  \textbf{\bibinfo{volume}{910}}, \bibinfo{pages}{L13}
  (\bibinfo{year}{2021}{\natexlab{b}}).

\bibitem[{\citenamefont{{The Event Horizon Telescope Collaboration, et
  al.}}(2019)}]{Collaboration_2019}
\bibinfo{author}{\bibnamefont{{The Event Horizon Telescope Collaboration, et
  al.}}}, \bibinfo{journal}{Astrophys. J. Lett.}
  \textbf{\bibinfo{volume}{875}}, \bibinfo{pages}{L1} (\bibinfo{year}{2019}).

\bibitem[{\citenamefont{Einstein}(2015)}]{Einstein_1916}
\bibinfo{author}{\bibfnamefont{A.}~\bibnamefont{Einstein}},
  \emph{\bibinfo{title}{English Translation of “The Foundation of the General
  Theory of Relativity”}} (\bibinfo{publisher}{Princeton University Press},
  \bibinfo{year}{2015}).

\bibitem[{\citenamefont{Ozel et~al.}(2021)\citenamefont{Ozel, Psaltis, and
  Younsi}}]{Ozel:2021ayr}
\bibinfo{author}{\bibfnamefont{F.}~\bibnamefont{Ozel}},
  \bibinfo{author}{\bibfnamefont{D.}~\bibnamefont{Psaltis}}, \bibnamefont{and}
  \bibinfo{author}{\bibfnamefont{Z.}~\bibnamefont{Younsi}}
  (\bibinfo{year}{2021}), \eprint{arXiv: 2111.01123}.

\bibitem[{\citenamefont{Vagnozzi et~al.}(2020)\citenamefont{Vagnozzi, Bambi,
  and Visinelli}}]{Vagnozzi:2020quf}
\bibinfo{author}{\bibfnamefont{S.}~\bibnamefont{Vagnozzi}},
  \bibinfo{author}{\bibfnamefont{C.}~\bibnamefont{Bambi}}, \bibnamefont{and}
  \bibinfo{author}{\bibfnamefont{L.}~\bibnamefont{Visinelli}},
  \bibinfo{journal}{Class. Quant. Grav.} \textbf{\bibinfo{volume}{37}},
  \bibinfo{pages}{087001} (\bibinfo{year}{2020}).

\bibitem[{\citenamefont{Allahyari et~al.}(2020)\citenamefont{Allahyari,
  Khodadi, Vagnozzi, and Mota}}]{Allahyari:2019jqz}
\bibinfo{author}{\bibfnamefont{A.}~\bibnamefont{Allahyari}},
  \bibinfo{author}{\bibfnamefont{M.}~\bibnamefont{Khodadi}},
  \bibinfo{author}{\bibfnamefont{S.}~\bibnamefont{Vagnozzi}}, \bibnamefont{and}
  \bibinfo{author}{\bibfnamefont{D.~F.} \bibnamefont{Mota}},
  \bibinfo{journal}{JCAP} \textbf{\bibinfo{volume}{02}}, \bibinfo{pages}{003}
  (\bibinfo{year}{2020}).

\bibitem[{\citenamefont{Vagnozzi and Visinelli}(2019)}]{Vagnozzi:2019apd}
\bibinfo{author}{\bibfnamefont{S.}~\bibnamefont{Vagnozzi}} \bibnamefont{and}
  \bibinfo{author}{\bibfnamefont{L.}~\bibnamefont{Visinelli}},
  \bibinfo{journal}{Phys. Rev. D} \textbf{\bibinfo{volume}{100}},
  \bibinfo{pages}{024020} (\bibinfo{year}{2019}).

\bibitem[{\citenamefont{Guerrero et~al.}(2021)\citenamefont{Guerrero, Olmo,
  Rubiera-Garcia, and G\'omez}}]{Guerrero:2021ues}
\bibinfo{author}{\bibfnamefont{M.}~\bibnamefont{Guerrero}},
  \bibinfo{author}{\bibfnamefont{G.~J.} \bibnamefont{Olmo}},
  \bibinfo{author}{\bibfnamefont{D.}~\bibnamefont{Rubiera-Garcia}},
  \bibnamefont{and} \bibinfo{author}{\bibfnamefont{D.~S.-C.}
  \bibnamefont{G\'omez}}, \bibinfo{journal}{JCAP}
  \textbf{\bibinfo{volume}{08}}, \bibinfo{pages}{036} (\bibinfo{year}{2021}).

\bibitem[{\citenamefont{Berti et~al.}(2015)}]{Berti:2015itd}
\bibinfo{author}{\bibfnamefont{E.}~\bibnamefont{Berti}} \bibnamefont{et~al.},
  \bibinfo{journal}{Class. Quant. Grav.} \textbf{\bibinfo{volume}{32}},
  \bibinfo{pages}{243001} (\bibinfo{year}{2015}).

\bibitem[{\citenamefont{Cunha and Herdeiro}(2018)}]{Cunha:2018acu}
\bibinfo{author}{\bibfnamefont{P.~V.~P.} \bibnamefont{Cunha}} \bibnamefont{and}
  \bibinfo{author}{\bibfnamefont{C.~A.~R.} \bibnamefont{Herdeiro}},
  \bibinfo{journal}{Gen. Rel. Grav.} \textbf{\bibinfo{volume}{50}},
  \bibinfo{pages}{42} (\bibinfo{year}{2018}).

\bibitem[{\citenamefont{Barausse et~al.}(2020)}]{Barausse:2020rsu}
\bibinfo{author}{\bibfnamefont{E.}~\bibnamefont{Barausse}}
  \bibnamefont{et~al.}, \bibinfo{journal}{Gen. Rel. Grav.}
  \textbf{\bibinfo{volume}{52}}, \bibinfo{pages}{81} (\bibinfo{year}{2020}).

\bibitem[{\citenamefont{Cunha et~al.}(2017)\citenamefont{Cunha, Berti, and
  Herdeiro}}]{Cunha:2017qtt}
\bibinfo{author}{\bibfnamefont{P.~V.~P.} \bibnamefont{Cunha}},
  \bibinfo{author}{\bibfnamefont{E.}~\bibnamefont{Berti}}, \bibnamefont{and}
  \bibinfo{author}{\bibfnamefont{C.~A.~R.} \bibnamefont{Herdeiro}},
  \bibinfo{journal}{Phys. Rev. Lett.} \textbf{\bibinfo{volume}{119}},
  \bibinfo{pages}{251102} (\bibinfo{year}{2017}).

\bibitem[{\citenamefont{Junior et~al.}(2021)\citenamefont{Junior, Cunha,
  Herdeiro, and Crispino}}]{Junior:2021dyw}
\bibinfo{author}{\bibfnamefont{H.~C. D.~L.} \bibnamefont{Junior}},
  \bibinfo{author}{\bibfnamefont{P.~V.~P.} \bibnamefont{Cunha}},
  \bibinfo{author}{\bibfnamefont{C.~A.~R.} \bibnamefont{Herdeiro}},
  \bibnamefont{and} \bibinfo{author}{\bibfnamefont{L.~C.~B.}
  \bibnamefont{Crispino}}, \bibinfo{journal}{Phys. Rev. D}
  \textbf{\bibinfo{volume}{104}}, \bibinfo{pages}{044018}
  (\bibinfo{year}{2021}).

\bibitem[{\citenamefont{Cunha et~al.}(2020)\citenamefont{Cunha, Eir\'o,
  Herdeiro, and Lemos}}]{Cunha:2019hzj}
\bibinfo{author}{\bibfnamefont{P.~V.~P.} \bibnamefont{Cunha}},
  \bibinfo{author}{\bibfnamefont{N.~A.} \bibnamefont{Eir\'o}},
  \bibinfo{author}{\bibfnamefont{C.~A.~R.} \bibnamefont{Herdeiro}},
  \bibnamefont{and} \bibinfo{author}{\bibfnamefont{J.~P.~S.}
  \bibnamefont{Lemos}}, \bibinfo{journal}{JCAP} \textbf{\bibinfo{volume}{03}},
  \bibinfo{pages}{035} (\bibinfo{year}{2020}).

\bibitem[{\citenamefont{Abdujabbarov et~al.}(2017)\citenamefont{Abdujabbarov,
  Ahmedov, Dadhich, and Atamurotov}}]{Abdujabbarov:2017pfw}
\bibinfo{author}{\bibfnamefont{A.}~\bibnamefont{Abdujabbarov}},
  \bibinfo{author}{\bibfnamefont{B.}~\bibnamefont{Ahmedov}},
  \bibinfo{author}{\bibfnamefont{N.}~\bibnamefont{Dadhich}}, \bibnamefont{and}
  \bibinfo{author}{\bibfnamefont{F.}~\bibnamefont{Atamurotov}},
  \bibinfo{journal}{Phys. Rev. D} \textbf{\bibinfo{volume}{96}},
  \bibinfo{pages}{084017} (\bibinfo{year}{2017}).

\bibitem[{\citenamefont{Narzilloev et~al.}(2021)\citenamefont{Narzilloev,
  Shaymatov, Hussain, Abdujabbarov, Ahmedov, and Bambi}}]{Narzilloev:2021jtg}
\bibinfo{author}{\bibfnamefont{B.}~\bibnamefont{Narzilloev}},
  \bibinfo{author}{\bibfnamefont{S.}~\bibnamefont{Shaymatov}},
  \bibinfo{author}{\bibfnamefont{I.}~\bibnamefont{Hussain}},
  \bibinfo{author}{\bibfnamefont{A.}~\bibnamefont{Abdujabbarov}},
  \bibinfo{author}{\bibfnamefont{B.}~\bibnamefont{Ahmedov}}, \bibnamefont{and}
  \bibinfo{author}{\bibfnamefont{C.}~\bibnamefont{Bambi}},
  \bibinfo{journal}{Eur. Phys. J. C} \textbf{\bibinfo{volume}{81}},
  \bibinfo{pages}{849} (\bibinfo{year}{2021}).

\bibitem[{\citenamefont{Jarosik et~al.}(2011)\citenamefont{Jarosik, Bennett,
  Dunkley et~al.}}]{Jarosik2011}
\bibinfo{author}{\bibfnamefont{N.}~\bibnamefont{Jarosik}},
  \bibinfo{author}{\bibfnamefont{C.~L.} \bibnamefont{Bennett}},
  \bibinfo{author}{\bibfnamefont{J.}~\bibnamefont{Dunkley}},
  \bibnamefont{et~al.}, \bibinfo{journal}{Astrophys. J., Suppl. Ser.}
  \textbf{\bibinfo{volume}{192}}, \bibinfo{pages}{14} (\bibinfo{year}{2011}).

\bibitem[{\citenamefont{Bernabei et~al.}(2008)\citenamefont{Bernabei, Belli,
  Cappella et~al.}}]{Bernabei2008}
\bibinfo{author}{\bibfnamefont{R.}~\bibnamefont{Bernabei}},
  \bibinfo{author}{\bibfnamefont{P.}~\bibnamefont{Belli}},
  \bibinfo{author}{\bibfnamefont{F.}~\bibnamefont{Cappella}},
  \bibnamefont{et~al.}, \bibinfo{journal}{Eur. Phys. J. C}
  \textbf{\bibinfo{volume}{56}}, \bibinfo{pages}{333} (\bibinfo{year}{2008}).

\bibitem[{\citenamefont{Bernabei et~al.}(2013)\citenamefont{Bernabei, Belli,
  Cappella et~al.}}]{Bernabei2013}
\bibinfo{author}{\bibfnamefont{R.}~\bibnamefont{Bernabei}},
  \bibinfo{author}{\bibfnamefont{P.}~\bibnamefont{Belli}},
  \bibinfo{author}{\bibfnamefont{F.}~\bibnamefont{Cappella}},
  \bibnamefont{et~al.}, \bibinfo{journal}{Eur. Phys. J. C}
  \textbf{\bibinfo{volume}{73}}, \bibinfo{pages}{2648} (\bibinfo{year}{2013}).

\bibitem[{\citenamefont{Bernabei et~al.}(2018)\citenamefont{Bernabei, Belli,
  Bussolotti et~al.}}]{Bernabei2018}
\bibinfo{author}{\bibfnamefont{R.}~\bibnamefont{Bernabei}},
  \bibinfo{author}{\bibfnamefont{P.}~\bibnamefont{Belli}},
  \bibinfo{author}{\bibfnamefont{A.}~\bibnamefont{Bussolotti}},
  \bibnamefont{et~al.}, \bibinfo{journal}{Nucl. Phys. At. Energy}
  \textbf{\bibinfo{volume}{19}}, \bibinfo{pages}{307} (\bibinfo{year}{2018}).

\bibitem[{\citenamefont{Angloher et~al.}(2016)\citenamefont{Angloher, Bento,
  Bucci et~al.}}]{Angholer2016}
\bibinfo{author}{\bibfnamefont{G.}~\bibnamefont{Angloher}},
  \bibinfo{author}{\bibfnamefont{A.}~\bibnamefont{Bento}},
  \bibinfo{author}{\bibfnamefont{C.}~\bibnamefont{Bucci}},
  \bibnamefont{et~al.}, \bibinfo{journal}{Eur. Phys. J. C}
  \textbf{\bibinfo{volume}{76}}, \bibinfo{pages}{25} (\bibinfo{year}{2016}).

\bibitem[{\citenamefont{Amole et~al.}(2017)\citenamefont{Amole, Ardid,
  Arnquist, Asner et~al.}}]{Amole2017}
\bibinfo{author}{\bibfnamefont{C.}~\bibnamefont{Amole}},
  \bibinfo{author}{\bibfnamefont{M.}~\bibnamefont{Ardid}},
  \bibinfo{author}{\bibfnamefont{I.~J.} \bibnamefont{Arnquist}},
  \bibinfo{author}{\bibfnamefont{D.~M.} \bibnamefont{Asner}},
  \bibnamefont{et~al.} (\bibinfo{collaboration}{PICO Collaboration}),
  \bibinfo{journal}{Phys. Rev. Lett.} \textbf{\bibinfo{volume}{118}},
  \bibinfo{pages}{251301} (\bibinfo{year}{2017}).

\bibitem[{\citenamefont{Akerib et~al.}(2017)\citenamefont{Akerib, Alsum,
  Ara\'ujo, Bai et~al.}}]{Akerib2017}
\bibinfo{author}{\bibfnamefont{D.~S.} \bibnamefont{Akerib}},
  \bibinfo{author}{\bibfnamefont{S.}~\bibnamefont{Alsum}},
  \bibinfo{author}{\bibfnamefont{H.~M.} \bibnamefont{Ara\'ujo}},
  \bibinfo{author}{\bibfnamefont{X.}~\bibnamefont{Bai}}, \bibnamefont{et~al.}
  (\bibinfo{collaboration}{LUX Collaboration}), \bibinfo{journal}{Phys. Rev.
  Lett.} \textbf{\bibinfo{volume}{118}}, \bibinfo{pages}{251302}
  (\bibinfo{year}{2017}).

\bibitem[{\citenamefont{Baum et~al.}(2020)\citenamefont{Baum, Drukier, Freese
  et~al.}}]{Baum2020}
\bibinfo{author}{\bibfnamefont{S.}~\bibnamefont{Baum}},
  \bibinfo{author}{\bibfnamefont{A.~K.} \bibnamefont{Drukier}},
  \bibinfo{author}{\bibfnamefont{K.}~\bibnamefont{Freese}},
  \bibnamefont{et~al.}, \bibinfo{journal}{Phys. Lett. B}
  \textbf{\bibinfo{volume}{803}}, \bibinfo{pages}{135235}
  (\bibinfo{year}{2020}).

\bibitem[{\citenamefont{Hou et~al.}(2018{\natexlab{a}})\citenamefont{Hou, Xu,
  and Wang}}]{Hou_2018b}
\bibinfo{author}{\bibfnamefont{X.}~\bibnamefont{Hou}},
  \bibinfo{author}{\bibfnamefont{Z.}~\bibnamefont{Xu}}, \bibnamefont{and}
  \bibinfo{author}{\bibfnamefont{J.}~\bibnamefont{Wang}}, \bibinfo{journal}{J.
  Cosmol. Astropart. Phys.} \textbf{\bibinfo{volume}{2018}},
  \bibinfo{pages}{040} (\bibinfo{year}{2018}{\natexlab{a}}).

\bibitem[{\citenamefont{Konoplya}(2019)}]{Konoplya_2019}
\bibinfo{author}{\bibfnamefont{R.~A.} \bibnamefont{Konoplya}},
  \bibinfo{journal}{Phys. Lett. B} \textbf{\bibinfo{volume}{795}},
  \bibinfo{pages}{1} (\bibinfo{year}{2019}).

\bibitem[{\citenamefont{Pantig and Rodulfo}(2020{\natexlab{a}})}]{Pantig2020b}
\bibinfo{author}{\bibfnamefont{R.~C.} \bibnamefont{Pantig}} \bibnamefont{and}
  \bibinfo{author}{\bibfnamefont{E.~T.} \bibnamefont{Rodulfo}},
  \bibinfo{journal}{Chinese J. Phys.} \textbf{\bibinfo{volume}{68}},
  \bibinfo{pages}{236} (\bibinfo{year}{2020}{\natexlab{a}}).

\bibitem[{\citenamefont{Pantig and
  Rodulfo}(2020{\natexlab{b}})}]{Pantig:2020odu}
\bibinfo{author}{\bibfnamefont{R.~C.} \bibnamefont{Pantig}} \bibnamefont{and}
  \bibinfo{author}{\bibfnamefont{E.~T.} \bibnamefont{Rodulfo}},
  \bibinfo{journal}{Chin. J. Phys.} \textbf{\bibinfo{volume}{66}},
  \bibinfo{pages}{691} (\bibinfo{year}{2020}{\natexlab{b}}).

\bibitem[{\citenamefont{Pantig et~al.}(2022)\citenamefont{Pantig, Yu, Rodulfo,
  and \"Ovg\"un}}]{Pantig2022}
\bibinfo{author}{\bibfnamefont{R.~C.} \bibnamefont{Pantig}},
  \bibinfo{author}{\bibfnamefont{P.~K.} \bibnamefont{Yu}},
  \bibinfo{author}{\bibfnamefont{E.~T.} \bibnamefont{Rodulfo}},
  \bibnamefont{and}
  \bibinfo{author}{\bibfnamefont{A.}~\bibnamefont{\"Ovg\"un}},
  \bibinfo{journal}{Annals of Physics} \textbf{\bibinfo{volume}{436}},
  \bibinfo{pages}{168722} (\bibinfo{year}{2022}).

\bibitem[{\citenamefont{Saurabh and Jusufi}(2021)}]{Saurabh2021}
\bibinfo{author}{\bibfnamefont{K.}~\bibnamefont{Saurabh}} \bibnamefont{and}
  \bibinfo{author}{\bibfnamefont{K.}~\bibnamefont{Jusufi}},
  \bibinfo{journal}{Eur. Phys. J. C} \textbf{\bibinfo{volume}{81}},
  \bibinfo{pages}{490} (\bibinfo{year}{2021}).

\bibitem[{\citenamefont{Xu et~al.}(2018)\citenamefont{Xu, Hou, Gong, and
  Wang}}]{Xu_2018}
\bibinfo{author}{\bibfnamefont{Z.}~\bibnamefont{Xu}},
  \bibinfo{author}{\bibfnamefont{X.}~\bibnamefont{Hou}},
  \bibinfo{author}{\bibfnamefont{X.}~\bibnamefont{Gong}}, \bibnamefont{and}
  \bibinfo{author}{\bibfnamefont{J.}~\bibnamefont{Wang}},
  \bibinfo{journal}{JCAP} \textbf{\bibinfo{volume}{09}}, \bibinfo{pages}{038}
  (\bibinfo{year}{2018}).

\bibitem[{\citenamefont{Gibbons and Werner}(2008)}]{Gibbons_2008}
\bibinfo{author}{\bibfnamefont{G.~W.} \bibnamefont{Gibbons}} \bibnamefont{and}
  \bibinfo{author}{\bibfnamefont{M.~C.} \bibnamefont{Werner}},
  \bibinfo{journal}{Class. Quantum Gravity} \textbf{\bibinfo{volume}{25}},
  \bibinfo{pages}{235009} (\bibinfo{year}{2008}).

\bibitem[{\citenamefont{Werner}(2012{\natexlab{a}})}]{Werner_2012}
\bibinfo{author}{\bibfnamefont{M.~C.} \bibnamefont{Werner}},
  \bibinfo{journal}{Gen. Relativ. Gravit.} \textbf{\bibinfo{volume}{44}},
  \bibinfo{pages}{3047} (\bibinfo{year}{2012}{\natexlab{a}}).

\bibitem[{\citenamefont{Ishihara et~al.}(2016)\citenamefont{Ishihara, Suzuki,
  Ono et~al.}}]{Ishihara_2016}
\bibinfo{author}{\bibfnamefont{A.}~\bibnamefont{Ishihara}},
  \bibinfo{author}{\bibfnamefont{Y.}~\bibnamefont{Suzuki}},
  \bibinfo{author}{\bibfnamefont{T.}~\bibnamefont{Ono}}, \bibnamefont{et~al.},
  \bibinfo{journal}{Phys. Rev. D} \textbf{\bibinfo{volume}{94}},
  \bibinfo{pages}{084015} (\bibinfo{year}{2016}).

\bibitem[{\citenamefont{Ishihara et~al.}(2017)\citenamefont{Ishihara, Suzuki,
  Ono, and Asada}}]{Ishihara:2016sfv}
\bibinfo{author}{\bibfnamefont{A.}~\bibnamefont{Ishihara}},
  \bibinfo{author}{\bibfnamefont{Y.}~\bibnamefont{Suzuki}},
  \bibinfo{author}{\bibfnamefont{T.}~\bibnamefont{Ono}}, \bibnamefont{and}
  \bibinfo{author}{\bibfnamefont{H.}~\bibnamefont{Asada}},
  \bibinfo{journal}{Phys. Rev. D} \textbf{\bibinfo{volume}{95}},
  \bibinfo{pages}{044017} (\bibinfo{year}{2017}).

\bibitem[{\citenamefont{Ono et~al.}(2017)\citenamefont{Ono, Ishihara, and
  Asada}}]{Ono:2017pie}
\bibinfo{author}{\bibfnamefont{T.}~\bibnamefont{Ono}},
  \bibinfo{author}{\bibfnamefont{A.}~\bibnamefont{Ishihara}}, \bibnamefont{and}
  \bibinfo{author}{\bibfnamefont{H.}~\bibnamefont{Asada}},
  \bibinfo{journal}{Phys. Rev. D} \textbf{\bibinfo{volume}{96}},
  \bibinfo{pages}{104037} (\bibinfo{year}{2017}).

\bibitem[{\citenamefont{Crisnejo and Gallo}(2018)}]{Crisnejo:2018uyn}
\bibinfo{author}{\bibfnamefont{G.}~\bibnamefont{Crisnejo}} \bibnamefont{and}
  \bibinfo{author}{\bibfnamefont{E.}~\bibnamefont{Gallo}},
  \bibinfo{journal}{Phys. Rev. D} \textbf{\bibinfo{volume}{97}},
  \bibinfo{pages}{124016} (\bibinfo{year}{2018}).

\bibitem[{\citenamefont{Li and \"Ovg\"un}(2020)}]{Li:2020dln}
\bibinfo{author}{\bibfnamefont{Z.}~\bibnamefont{Li}} \bibnamefont{and}
  \bibinfo{author}{\bibfnamefont{A.}~\bibnamefont{\"Ovg\"un}},
  \bibinfo{journal}{Phys. Rev. D} \textbf{\bibinfo{volume}{101}},
  \bibinfo{pages}{024040} (\bibinfo{year}{2020}).

\bibitem[{\citenamefont{Li et~al.}(2020)\citenamefont{Li, Zhang, and
  \"Ovg\"un}}]{Li:2020wvn}
\bibinfo{author}{\bibfnamefont{Z.}~\bibnamefont{Li}},
  \bibinfo{author}{\bibfnamefont{G.}~\bibnamefont{Zhang}}, \bibnamefont{and}
  \bibinfo{author}{\bibfnamefont{A.}~\bibnamefont{\"Ovg\"un}},
  \bibinfo{journal}{Phys. Rev. D} \textbf{\bibinfo{volume}{101}},
  \bibinfo{pages}{124058} (\bibinfo{year}{2020}).

\bibitem[{\citenamefont{Werner}(2012{\natexlab{b}})}]{Werner:2012rc}
\bibinfo{author}{\bibfnamefont{M.~C.} \bibnamefont{Werner}},
  \bibinfo{journal}{Gen. Rel. Grav.} \textbf{\bibinfo{volume}{44}},
  \bibinfo{pages}{3047} (\bibinfo{year}{2012}{\natexlab{b}}).

\bibitem[{\citenamefont{Gibbons et~al.}(2009)\citenamefont{Gibbons, Herdeiro,
  Warnick, and Werner}}]{Gibbons:2008zi}
\bibinfo{author}{\bibfnamefont{G.~W.} \bibnamefont{Gibbons}},
  \bibinfo{author}{\bibfnamefont{C.~A.~R.} \bibnamefont{Herdeiro}},
  \bibinfo{author}{\bibfnamefont{C.~M.} \bibnamefont{Warnick}},
  \bibnamefont{and} \bibinfo{author}{\bibfnamefont{M.~C.}
  \bibnamefont{Werner}}, \bibinfo{journal}{Phys. Rev. D}
  \textbf{\bibinfo{volume}{79}}, \bibinfo{pages}{044022}
  (\bibinfo{year}{2009}).

\bibitem[{\citenamefont{Jusufi and
  \"Ovg\"un}(2018{\natexlab{a}})}]{Jusufi:2017mav}
\bibinfo{author}{\bibfnamefont{K.}~\bibnamefont{Jusufi}} \bibnamefont{and}
  \bibinfo{author}{\bibfnamefont{A.}~\bibnamefont{\"Ovg\"un}},
  \bibinfo{journal}{Phys. Rev. D} \textbf{\bibinfo{volume}{97}},
  \bibinfo{pages}{024042} (\bibinfo{year}{2018}{\natexlab{a}}).

\bibitem[{\citenamefont{Jusufi et~al.}(2017)\citenamefont{Jusufi, Werner,
  Banerjee, and \"Ovg\"un}}]{Jusufi:2017lsl}
\bibinfo{author}{\bibfnamefont{K.}~\bibnamefont{Jusufi}},
  \bibinfo{author}{\bibfnamefont{M.~C.} \bibnamefont{Werner}},
  \bibinfo{author}{\bibfnamefont{A.}~\bibnamefont{Banerjee}}, \bibnamefont{and}
  \bibinfo{author}{\bibfnamefont{A.}~\bibnamefont{\"Ovg\"un}},
  \bibinfo{journal}{Phys. Rev. D} \textbf{\bibinfo{volume}{95}},
  \bibinfo{pages}{104012} (\bibinfo{year}{2017}).

\bibitem[{\citenamefont{\"Ovg\"un}(2018)}]{Ovgun:2018fnk}
\bibinfo{author}{\bibfnamefont{A.}~\bibnamefont{\"Ovg\"un}},
  \bibinfo{journal}{Phys. Rev. D} \textbf{\bibinfo{volume}{98}},
  \bibinfo{pages}{044033} (\bibinfo{year}{2018}).

\bibitem[{\citenamefont{Ono et~al.}(2018)\citenamefont{Ono, Ishihara, and
  Asada}}]{Ono:2018ybw}
\bibinfo{author}{\bibfnamefont{T.}~\bibnamefont{Ono}},
  \bibinfo{author}{\bibfnamefont{A.}~\bibnamefont{Ishihara}}, \bibnamefont{and}
  \bibinfo{author}{\bibfnamefont{H.}~\bibnamefont{Asada}},
  \bibinfo{journal}{Phys. Rev. D} \textbf{\bibinfo{volume}{98}},
  \bibinfo{pages}{044047} (\bibinfo{year}{2018}).

\bibitem[{\citenamefont{Jusufi and
  \"Ovg\"un}(2018{\natexlab{b}})}]{Jusufi:2017uhh}
\bibinfo{author}{\bibfnamefont{K.}~\bibnamefont{Jusufi}} \bibnamefont{and}
  \bibinfo{author}{\bibfnamefont{A.}~\bibnamefont{\"Ovg\"un}},
  \bibinfo{journal}{Phys. Rev. D} \textbf{\bibinfo{volume}{97}},
  \bibinfo{pages}{064030} (\bibinfo{year}{2018}{\natexlab{b}}).

\bibitem[{\citenamefont{Javed et~al.}(2019{\natexlab{a}})\citenamefont{Javed,
  Babar, and \"Ovg\"un}}]{Javed:2019qyg}
\bibinfo{author}{\bibfnamefont{W.}~\bibnamefont{Javed}},
  \bibinfo{author}{\bibfnamefont{R.}~\bibnamefont{Babar}}, \bibnamefont{and}
  \bibinfo{author}{\bibfnamefont{A.}~\bibnamefont{\"Ovg\"un}},
  \bibinfo{journal}{Phys. Rev. D} \textbf{\bibinfo{volume}{99}},
  \bibinfo{pages}{084012} (\bibinfo{year}{2019}{\natexlab{a}}).

\bibitem[{\citenamefont{Arakida}(2018)}]{Arakida:2017hrm}
\bibinfo{author}{\bibfnamefont{H.}~\bibnamefont{Arakida}},
  \bibinfo{journal}{Gen. Rel. Grav.} \textbf{\bibinfo{volume}{50}},
  \bibinfo{pages}{48} (\bibinfo{year}{2018}).

\bibitem[{\citenamefont{\"Ovg\"un}(2019{\natexlab{a}})}]{Ovgun:2019wej}
\bibinfo{author}{\bibfnamefont{A.}~\bibnamefont{\"Ovg\"un}},
  \bibinfo{journal}{Phys. Rev. D} \textbf{\bibinfo{volume}{99}},
  \bibinfo{pages}{104075} (\bibinfo{year}{2019}{\natexlab{a}}).

\bibitem[{\citenamefont{Gibbons}(2016)}]{Gibbons:2015qja}
\bibinfo{author}{\bibfnamefont{G.~W.} \bibnamefont{Gibbons}},
  \bibinfo{journal}{Class. Quant. Grav.} \textbf{\bibinfo{volume}{33}},
  \bibinfo{pages}{025004} (\bibinfo{year}{2016}).

\bibitem[{\citenamefont{\"Ovg\"un et~al.}(2019)\citenamefont{\"Ovg\"un,
  Gyulchev, and Jusufi}}]{Ovgun:2018prw}
\bibinfo{author}{\bibfnamefont{A.}~\bibnamefont{\"Ovg\"un}},
  \bibinfo{author}{\bibfnamefont{G.}~\bibnamefont{Gyulchev}}, \bibnamefont{and}
  \bibinfo{author}{\bibfnamefont{K.}~\bibnamefont{Jusufi}},
  \bibinfo{journal}{Annals Phys.} \textbf{\bibinfo{volume}{406}},
  \bibinfo{pages}{152} (\bibinfo{year}{2019}).

\bibitem[{\citenamefont{Ono et~al.}(2019)\citenamefont{Ono, Ishihara, and
  Asada}}]{Ono:2018jrv}
\bibinfo{author}{\bibfnamefont{T.}~\bibnamefont{Ono}},
  \bibinfo{author}{\bibfnamefont{A.}~\bibnamefont{Ishihara}}, \bibnamefont{and}
  \bibinfo{author}{\bibfnamefont{H.}~\bibnamefont{Asada}},
  \bibinfo{journal}{Phys. Rev. D} \textbf{\bibinfo{volume}{99}},
  \bibinfo{pages}{124030} (\bibinfo{year}{2019}).

\bibitem[{\citenamefont{\"Ovg\"un}(2019{\natexlab{b}})}]{Ovgun:2018oxk}
\bibinfo{author}{\bibfnamefont{A.}~\bibnamefont{\"Ovg\"un}},
  \bibinfo{journal}{Universe} \textbf{\bibinfo{volume}{5}},
  \bibinfo{pages}{115} (\bibinfo{year}{2019}{\natexlab{b}}).

\bibitem[{\citenamefont{Javed et~al.}(2019{\natexlab{b}})\citenamefont{Javed,
  Abbas, and \"Ovg\"un}}]{Javed:2019rrg}
\bibinfo{author}{\bibfnamefont{W.}~\bibnamefont{Javed}},
  \bibinfo{author}{\bibfnamefont{j.}~\bibnamefont{Abbas}}, \bibnamefont{and}
  \bibinfo{author}{\bibfnamefont{A.}~\bibnamefont{\"Ovg\"un}},
  \bibinfo{journal}{Phys. Rev. D} \textbf{\bibinfo{volume}{100}},
  \bibinfo{pages}{044052} (\bibinfo{year}{2019}{\natexlab{b}}).

\bibitem[{\citenamefont{Crisnejo et~al.}(2019)\citenamefont{Crisnejo, Gallo,
  and Rogers}}]{Crisnejo:2018ppm}
\bibinfo{author}{\bibfnamefont{G.}~\bibnamefont{Crisnejo}},
  \bibinfo{author}{\bibfnamefont{E.}~\bibnamefont{Gallo}}, \bibnamefont{and}
  \bibinfo{author}{\bibfnamefont{A.}~\bibnamefont{Rogers}},
  \bibinfo{journal}{Phys. Rev. D} \textbf{\bibinfo{volume}{99}},
  \bibinfo{pages}{124001} (\bibinfo{year}{2019}).

\bibitem[{\citenamefont{Javed et~al.}(2019{\natexlab{c}})\citenamefont{Javed,
  Babar, and \"Ovg\"un}}]{Javed:2019ynm}
\bibinfo{author}{\bibfnamefont{W.}~\bibnamefont{Javed}},
  \bibinfo{author}{\bibfnamefont{R.}~\bibnamefont{Babar}}, \bibnamefont{and}
  \bibinfo{author}{\bibfnamefont{A.}~\bibnamefont{\"Ovg\"un}},
  \bibinfo{journal}{Phys. Rev. D} \textbf{\bibinfo{volume}{100}},
  \bibinfo{pages}{104032} (\bibinfo{year}{2019}{\natexlab{c}}).

\bibitem[{\citenamefont{Jusufi and \"Ovg\"un}(2019)}]{Jusufi:2017vew}
\bibinfo{author}{\bibfnamefont{K.}~\bibnamefont{Jusufi}} \bibnamefont{and}
  \bibinfo{author}{\bibfnamefont{A.}~\bibnamefont{\"Ovg\"un}},
  \bibinfo{journal}{Int. J. Geom. Meth. Mod. Phys.}
  \textbf{\bibinfo{volume}{16}}, \bibinfo{pages}{1950116}
  (\bibinfo{year}{2019}).

\bibitem[{\citenamefont{Belhaj et~al.}(2020)\citenamefont{Belhaj, Benali,
  El~Balali, El~Moumni, and Ennadifi}}]{Belhaj:2020rdb}
\bibinfo{author}{\bibfnamefont{A.}~\bibnamefont{Belhaj}},
  \bibinfo{author}{\bibfnamefont{M.}~\bibnamefont{Benali}},
  \bibinfo{author}{\bibfnamefont{A.}~\bibnamefont{El~Balali}},
  \bibinfo{author}{\bibfnamefont{H.}~\bibnamefont{El~Moumni}},
  \bibnamefont{and} \bibinfo{author}{\bibfnamefont{S.~E.}
  \bibnamefont{Ennadifi}}, \bibinfo{journal}{Class. Quant. Grav.}
  \textbf{\bibinfo{volume}{37}}, \bibinfo{pages}{215004}
  (\bibinfo{year}{2020}).

\bibitem[{\citenamefont{Ono and Asada}(2019)}]{Ono:2019hkw}
\bibinfo{author}{\bibfnamefont{T.}~\bibnamefont{Ono}} \bibnamefont{and}
  \bibinfo{author}{\bibfnamefont{H.}~\bibnamefont{Asada}},
  \bibinfo{journal}{Universe} \textbf{\bibinfo{volume}{5}},
  \bibinfo{pages}{218} (\bibinfo{year}{2019}).

\bibitem[{\citenamefont{\"Ovg\"un and Sakall\i{}}(2020)}]{Ovgun:2020gjz}
\bibinfo{author}{\bibfnamefont{A.}~\bibnamefont{\"Ovg\"un}} \bibnamefont{and}
  \bibinfo{author}{\bibfnamefont{I.}~\bibnamefont{Sakall\i{}}},
  \bibinfo{journal}{Class. Quant. Grav.} \textbf{\bibinfo{volume}{37}},
  \bibinfo{pages}{225003} (\bibinfo{year}{2020}).

\bibitem[{\citenamefont{Javed et~al.}(2020{\natexlab{a}})\citenamefont{Javed,
  Abbas, and \"Ovg\"un}}]{Javed:2019jag}
\bibinfo{author}{\bibfnamefont{W.}~\bibnamefont{Javed}},
  \bibinfo{author}{\bibfnamefont{J.}~\bibnamefont{Abbas}}, \bibnamefont{and}
  \bibinfo{author}{\bibfnamefont{A.}~\bibnamefont{\"Ovg\"un}},
  \bibinfo{journal}{Annals Phys.} \textbf{\bibinfo{volume}{418}},
  \bibinfo{pages}{168183} (\bibinfo{year}{2020}{\natexlab{a}}).

\bibitem[{\citenamefont{Javed et~al.}(2020{\natexlab{b}})\citenamefont{Javed,
  Hamza, and \"Ovg\"un}}]{Javed:2020lsg}
\bibinfo{author}{\bibfnamefont{W.}~\bibnamefont{Javed}},
  \bibinfo{author}{\bibfnamefont{A.}~\bibnamefont{Hamza}}, \bibnamefont{and}
  \bibinfo{author}{\bibfnamefont{A.}~\bibnamefont{\"Ovg\"un}},
  \bibinfo{journal}{Phys. Rev. D} \textbf{\bibinfo{volume}{101}},
  \bibinfo{pages}{103521} (\bibinfo{year}{2020}{\natexlab{b}}).

\bibitem[{\citenamefont{Takizawa
  et~al.}(2020{\natexlab{a}})\citenamefont{Takizawa, Ono, and
  Asada}}]{Takizawa:2020egm}
\bibinfo{author}{\bibfnamefont{K.}~\bibnamefont{Takizawa}},
  \bibinfo{author}{\bibfnamefont{T.}~\bibnamefont{Ono}}, \bibnamefont{and}
  \bibinfo{author}{\bibfnamefont{H.}~\bibnamefont{Asada}},
  \bibinfo{journal}{Phys. Rev. D} \textbf{\bibinfo{volume}{101}},
  \bibinfo{pages}{104032} (\bibinfo{year}{2020}{\natexlab{a}}).

\bibitem[{\citenamefont{Javed et~al.}(2020{\natexlab{c}})\citenamefont{Javed,
  Khadim, \"Ovg\"un, and Abbas}}]{Javed:2020fli}
\bibinfo{author}{\bibfnamefont{W.}~\bibnamefont{Javed}},
  \bibinfo{author}{\bibfnamefont{M.~B.} \bibnamefont{Khadim}},
  \bibinfo{author}{\bibfnamefont{A.}~\bibnamefont{\"Ovg\"un}},
  \bibnamefont{and} \bibinfo{author}{\bibfnamefont{J.}~\bibnamefont{Abbas}},
  \bibinfo{journal}{Eur. Phys. J. Plus} \textbf{\bibinfo{volume}{135}},
  \bibinfo{pages}{314} (\bibinfo{year}{2020}{\natexlab{c}}).

\bibitem[{\citenamefont{\"Ovg\"un}(2020)}]{Ovgun2020}
\bibinfo{author}{\bibfnamefont{A.}~\bibnamefont{\"Ovg\"un}},
  \bibinfo{journal}{Turk. J. Phys.} \textbf{\bibinfo{volume}{44}},
  \bibinfo{pages}{465} (\bibinfo{year}{2020}), \eprint{2011.04423}.

\bibitem[{\citenamefont{Okyay and \"Ovg\"un}(2022)}]{Okyay:2021nnh}
\bibinfo{author}{\bibfnamefont{M.}~\bibnamefont{Okyay}} \bibnamefont{and}
  \bibinfo{author}{\bibfnamefont{A.}~\bibnamefont{\"Ovg\"un}},
  \bibinfo{journal}{JCAP} \textbf{\bibinfo{volume}{01}}, \bibinfo{pages}{009}
  (\bibinfo{year}{2022}), \eprint{2108.07766}.

\bibitem[{\citenamefont{Javed et~al.}(2020{\natexlab{d}})\citenamefont{Javed,
  Khadim, and \"Ovg\"un}}]{Javed:2020frq}
\bibinfo{author}{\bibfnamefont{W.}~\bibnamefont{Javed}},
  \bibinfo{author}{\bibfnamefont{M.~B.} \bibnamefont{Khadim}},
  \bibnamefont{and}
  \bibinfo{author}{\bibfnamefont{A.}~\bibnamefont{\"Ovg\"un}},
  \bibinfo{journal}{Eur. Phys. J. Plus} \textbf{\bibinfo{volume}{135}},
  \bibinfo{pages}{595} (\bibinfo{year}{2020}{\natexlab{d}}).

\bibitem[{\citenamefont{Takizawa
  et~al.}(2020{\natexlab{b}})\citenamefont{Takizawa, Ono, and
  Asada}}]{Takizawa:2020dja}
\bibinfo{author}{\bibfnamefont{K.}~\bibnamefont{Takizawa}},
  \bibinfo{author}{\bibfnamefont{T.}~\bibnamefont{Ono}}, \bibnamefont{and}
  \bibinfo{author}{\bibfnamefont{H.}~\bibnamefont{Asada}},
  \bibinfo{journal}{Phys. Rev. D} \textbf{\bibinfo{volume}{102}},
  \bibinfo{pages}{064060} (\bibinfo{year}{2020}{\natexlab{b}}).

\bibitem[{\citenamefont{Fu et~al.}(2021)\citenamefont{Fu, Zhao, and
  Liu}}]{Fu:2021akc}
\bibinfo{author}{\bibfnamefont{Q.-M.} \bibnamefont{Fu}},
  \bibinfo{author}{\bibfnamefont{L.}~\bibnamefont{Zhao}}, \bibnamefont{and}
  \bibinfo{author}{\bibfnamefont{Y.-X.} \bibnamefont{Liu}},
  \bibinfo{journal}{Phys. Rev. D} \textbf{\bibinfo{volume}{104}},
  \bibinfo{pages}{024033} (\bibinfo{year}{2021}).

\bibitem[{\citenamefont{Arakida}(2021)}]{Arakida:2020xil}
\bibinfo{author}{\bibfnamefont{H.}~\bibnamefont{Arakida}},
  \bibinfo{journal}{JCAP} \textbf{\bibinfo{volume}{08}}, \bibinfo{pages}{028}
  (\bibinfo{year}{2021}).

\bibitem[{\citenamefont{Kumaran and \"Ovg\"un}(2021)}]{Kumaran:2021rgj}
\bibinfo{author}{\bibfnamefont{Y.}~\bibnamefont{Kumaran}} \bibnamefont{and}
  \bibinfo{author}{\bibfnamefont{A.}~\bibnamefont{\"Ovg\"un}},
  \bibinfo{journal}{Turk. J. Phys.} \textbf{\bibinfo{volume}{45}},
  \bibinfo{pages}{247} (\bibinfo{year}{2021}).

\bibitem[{\citenamefont{Zhang}(2022)}]{Zhang:2021ygh}
\bibinfo{author}{\bibfnamefont{Z.}~\bibnamefont{Zhang}},
  \bibinfo{journal}{Class. Quant. Grav.} \textbf{\bibinfo{volume}{39}},
  \bibinfo{pages}{015003} (\bibinfo{year}{2022}), \eprint{2112.04149}.

\bibitem[{\citenamefont{Hou et~al.}(2018{\natexlab{b}})\citenamefont{Hou, Xu,
  Zhou et~al.}}]{Hou_2018a}
\bibinfo{author}{\bibfnamefont{X.}~\bibnamefont{Hou}},
  \bibinfo{author}{\bibfnamefont{Z.}~\bibnamefont{Xu}},
  \bibinfo{author}{\bibfnamefont{M.}~\bibnamefont{Zhou}}, \bibnamefont{et~al.},
  \bibinfo{journal}{J. Cosmol. Astropart. Phys.}
  \textbf{\bibinfo{volume}{2018}}, \bibinfo{pages}{015}
  (\bibinfo{year}{2018}{\natexlab{b}}).

\bibitem[{\citenamefont{Jusufi et~al.}(2019)\citenamefont{Jusufi, Jamil,
  Salucci et~al.}}]{Jusufi_2019}
\bibinfo{author}{\bibfnamefont{K.}~\bibnamefont{Jusufi}},
  \bibinfo{author}{\bibfnamefont{M.}~\bibnamefont{Jamil}},
  \bibinfo{author}{\bibfnamefont{P.}~\bibnamefont{Salucci}},
  \bibnamefont{et~al.}, \bibinfo{journal}{Phys. Rev. D}
  \textbf{\bibinfo{volume}{100}}, \bibinfo{pages}{044012}
  (\bibinfo{year}{2019}).

\bibitem[{\citenamefont{Nampalliwar et~al.}(2021)\citenamefont{Nampalliwar,
  Jusufi et~al.}}]{Nampalliwar2021}
\bibinfo{author}{\bibfnamefont{S.}~\bibnamefont{Nampalliwar},
  \bibfnamefont{S.~Kumar}},
  \bibinfo{author}{\bibfnamefont{K.}~\bibnamefont{Jusufi}},
  \bibnamefont{et~al.}, \bibinfo{journal}{Astrophys. J.}
  \textbf{\bibinfo{volume}{916}}, \bibinfo{pages}{116} (\bibinfo{year}{2021}).

\bibitem[{\citenamefont{Navarro et~al.}(1996)\citenamefont{Navarro, Frenk, and
  White}}]{Navarro:1995iw}
\bibinfo{author}{\bibfnamefont{J.~F.} \bibnamefont{Navarro}},
  \bibinfo{author}{\bibfnamefont{C.~S.} \bibnamefont{Frenk}}, \bibnamefont{and}
  \bibinfo{author}{\bibfnamefont{S.~D.~M.} \bibnamefont{White}},
  \bibinfo{journal}{Astrophys. J.} \textbf{\bibinfo{volume}{462}},
  \bibinfo{pages}{563} (\bibinfo{year}{1996}).

\bibitem[{\citenamefont{Navarro et~al.}(1997)\citenamefont{Navarro, Frenk, and
  White}}]{Navarro:1996gj}
\bibinfo{author}{\bibfnamefont{J.~F.} \bibnamefont{Navarro}},
  \bibinfo{author}{\bibfnamefont{C.~S.} \bibnamefont{Frenk}}, \bibnamefont{and}
  \bibinfo{author}{\bibfnamefont{S.~D.~M.} \bibnamefont{White}},
  \bibinfo{journal}{Astrophys. J.} \textbf{\bibinfo{volume}{490}},
  \bibinfo{pages}{493} (\bibinfo{year}{1997}).

\bibitem[{\citenamefont{Dubinski and Carlberg}(1991)}]{Dubinski:1991bm}
\bibinfo{author}{\bibfnamefont{J.}~\bibnamefont{Dubinski}} \bibnamefont{and}
  \bibinfo{author}{\bibfnamefont{R.~G.} \bibnamefont{Carlberg}},
  \bibinfo{journal}{Astrophys. J.} \textbf{\bibinfo{volume}{378}},
  \bibinfo{pages}{496} (\bibinfo{year}{1991}).

\bibitem[{\citenamefont{Spergel and Steinhardt}(2000)}]{Spergel:1999mh}
\bibinfo{author}{\bibfnamefont{D.~N.} \bibnamefont{Spergel}} \bibnamefont{and}
  \bibinfo{author}{\bibfnamefont{P.~J.} \bibnamefont{Steinhardt}},
  \bibinfo{journal}{Phys. Rev. Lett.} \textbf{\bibinfo{volume}{84}},
  \bibinfo{pages}{3760} (\bibinfo{year}{2000}).

\bibitem[{\citenamefont{Boehmer and Harko}(2007)}]{Boehmer:2007um}
\bibinfo{author}{\bibfnamefont{C.~G.} \bibnamefont{Boehmer}} \bibnamefont{and}
  \bibinfo{author}{\bibfnamefont{T.}~\bibnamefont{Harko}},
  \bibinfo{journal}{JCAP} \textbf{\bibinfo{volume}{06}}, \bibinfo{pages}{025}
  (\bibinfo{year}{2007}).

\bibitem[{\citenamefont{Persic et~al.}(1996)\citenamefont{Persic, Salucci, and
  Stel}}]{Persic:1995ru}
\bibinfo{author}{\bibfnamefont{M.}~\bibnamefont{Persic}},
  \bibinfo{author}{\bibfnamefont{P.}~\bibnamefont{Salucci}}, \bibnamefont{and}
  \bibinfo{author}{\bibfnamefont{F.}~\bibnamefont{Stel}},
  \bibinfo{journal}{Mon. Not. Roy. Astron. Soc.}
  \textbf{\bibinfo{volume}{281}}, \bibinfo{pages}{27} (\bibinfo{year}{1996}).

\bibitem[{\citenamefont{Xu et~al.}(2020)\citenamefont{Xu, Gong, and
  Zhang}}]{Xu2021a}
\bibinfo{author}{\bibfnamefont{Z.}~\bibnamefont{Xu}},
  \bibinfo{author}{\bibfnamefont{X.}~\bibnamefont{Gong}}, \bibnamefont{and}
  \bibinfo{author}{\bibfnamefont{S.-N.} \bibnamefont{Zhang}},
  \bibinfo{journal}{Phys. Rev. D} \textbf{\bibinfo{volume}{101}},
  \bibinfo{pages}{024029} (\bibinfo{year}{2020}).

\bibitem[{\citenamefont{Xu et~al.}(2021)\citenamefont{Xu, Wang, and
  Tang}}]{Xu2021b}
\bibinfo{author}{\bibfnamefont{Z.}~\bibnamefont{Xu}},
  \bibinfo{author}{\bibfnamefont{J.}~\bibnamefont{Wang}}, \bibnamefont{and}
  \bibinfo{author}{\bibfnamefont{M.}~\bibnamefont{Tang}}, \bibinfo{journal}{J.
  Cosmol. Astropart. Phys.} \textbf{\bibinfo{volume}{2021}},
  \bibinfo{pages}{007} (\bibinfo{year}{2021}).

\bibitem[{\citenamefont{Kaiser and Squires}(1993)}]{Kaiser1993}
\bibinfo{author}{\bibfnamefont{N.}~\bibnamefont{Kaiser}} \bibnamefont{and}
  \bibinfo{author}{\bibfnamefont{G.}~\bibnamefont{Squires}},
  \bibinfo{journal}{Astrophys. J.} \textbf{\bibinfo{volume}{404}},
  \bibinfo{pages}{441} (\bibinfo{year}{1993}).

\bibitem[{\citenamefont{Metcalf and Madau}(2001)}]{Metcalf2001}
\bibinfo{author}{\bibfnamefont{R.~B.} \bibnamefont{Metcalf}} \bibnamefont{and}
  \bibinfo{author}{\bibfnamefont{P.}~\bibnamefont{Madau}},
  \bibinfo{journal}{Astrophys. J.} \textbf{\bibinfo{volume}{563}},
  \bibinfo{pages}{9} (\bibinfo{year}{2001}).

\bibitem[{\citenamefont{\"Ovg\"un}(2019{\natexlab{c}})}]{Ovgun2019}
\bibinfo{author}{\bibfnamefont{A.}~\bibnamefont{\"Ovg\"un}},
  \bibinfo{journal}{Universe} \textbf{\bibinfo{volume}{5}},
  \bibinfo{pages}{115} (\bibinfo{year}{2019}{\natexlab{c}}).

\bibitem[{\citenamefont{Atamurotov et~al.}(2022)\citenamefont{Atamurotov,
  Papnoi, and Jusufi}}]{Atamurutov2022}
\bibinfo{author}{\bibfnamefont{F.}~\bibnamefont{Atamurotov}},
  \bibinfo{author}{\bibfnamefont{U.}~\bibnamefont{Papnoi}}, \bibnamefont{and}
  \bibinfo{author}{\bibfnamefont{K.}~\bibnamefont{Jusufi}},
  \bibinfo{journal}{Class. Quantum Gravity} \textbf{\bibinfo{volume}{39}},
  \bibinfo{pages}{025014} (\bibinfo{year}{2022}).

\bibitem[{\citenamefont{Do~Carmo}(2016)}]{Carmo2016}
\bibinfo{author}{\bibfnamefont{M.~P.} \bibnamefont{Do~Carmo}},
  \emph{\bibinfo{title}{Differential geometry of curves and surfaces: revised
  and updated second edition}} (\bibinfo{publisher}{Courier Dover
  Publications}, \bibinfo{year}{2016}).

\bibitem[{\citenamefont{Klingenberg}(2013)}]{Klingenberg2013}
\bibinfo{author}{\bibfnamefont{W.}~\bibnamefont{Klingenberg}},
  \emph{\bibinfo{title}{A course in differential geometry}},
  vol.~\bibinfo{volume}{51} (\bibinfo{publisher}{Springer Science \& Business
  Media}, \bibinfo{year}{2013}).

\bibitem[{\citenamefont{de~Blok}(2010)}]{deBlok_2010}
\bibinfo{author}{\bibfnamefont{W.}~\bibnamefont{de~Blok}},
  \bibinfo{journal}{Adv. Astron} \textbf{\bibinfo{volume}{2010}},
  \bibinfo{pages}{789293} (\bibinfo{year}{2010}).

\bibitem[{\citenamefont{Castignani et~al.}(2012)\citenamefont{Castignani,
  Frusciante, Vernieri, and Salucci}}]{Castignani:2012sr}
\bibinfo{author}{\bibfnamefont{G.}~\bibnamefont{Castignani}},
  \bibinfo{author}{\bibfnamefont{N.}~\bibnamefont{Frusciante}},
  \bibinfo{author}{\bibfnamefont{D.}~\bibnamefont{Vernieri}}, \bibnamefont{and}
  \bibinfo{author}{\bibfnamefont{P.}~\bibnamefont{Salucci}},
  \bibinfo{journal}{Natural Sci.} \textbf{\bibinfo{volume}{4}},
  \bibinfo{pages}{265} (\bibinfo{year}{2012}).

\bibitem[{\citenamefont{Salucci et~al.}(2007)\citenamefont{Salucci, Lapi,
  Tonini, Gentile, Yegorova, and Klein}}]{Salucci:2007tm}
\bibinfo{author}{\bibfnamefont{P.}~\bibnamefont{Salucci}},
  \bibinfo{author}{\bibfnamefont{A.}~\bibnamefont{Lapi}},
  \bibinfo{author}{\bibfnamefont{C.}~\bibnamefont{Tonini}},
  \bibinfo{author}{\bibfnamefont{G.}~\bibnamefont{Gentile}},
  \bibinfo{author}{\bibfnamefont{I.}~\bibnamefont{Yegorova}}, \bibnamefont{and}
  \bibinfo{author}{\bibfnamefont{U.}~\bibnamefont{Klein}},
  \bibinfo{journal}{Mon. Not. Roy. Astron. Soc.}
  \textbf{\bibinfo{volume}{378}}, \bibinfo{pages}{41} (\bibinfo{year}{2007}),
  \eprint{astro-ph/0703115}.

\bibitem[{\citenamefont{Nakashi et~al.}(2019)\citenamefont{Nakashi, Kobayashi,
  Ueda, and Saida}}]{Nakashi2019}
\bibinfo{author}{\bibfnamefont{K.}~\bibnamefont{Nakashi}},
  \bibinfo{author}{\bibfnamefont{S.}~\bibnamefont{Kobayashi}},
  \bibinfo{author}{\bibfnamefont{S.}~\bibnamefont{Ueda}}, \bibnamefont{and}
  \bibinfo{author}{\bibfnamefont{H.}~\bibnamefont{Saida}},
  \bibinfo{journal}{Prog. Theor. Exp. Phys. 2019}
  \textbf{\bibinfo{volume}{2019}}, \bibinfo{pages}{073E02}
  (\bibinfo{year}{2019}).

\bibitem[{\citenamefont{Lin and Li}(2019)}]{Lin2019}
\bibinfo{author}{\bibfnamefont{H.-N.} \bibnamefont{Lin}} \bibnamefont{and}
  \bibinfo{author}{\bibfnamefont{X.}~\bibnamefont{Li}}, \bibinfo{journal}{Mon.
  Not. R. Astron. Soc.} \textbf{\bibinfo{volume}{487}}, \bibinfo{pages}{5679}
  (\bibinfo{year}{2019}).

\bibitem[{\citenamefont{Bozza}(2008)}]{Bozza2008}
\bibinfo{author}{\bibfnamefont{V.}~\bibnamefont{Bozza}},
  \bibinfo{journal}{Phys. Rev. D} \textbf{\bibinfo{volume}{78}},
  \bibinfo{pages}{103005} (\bibinfo{year}{2008}).

\bibitem[{\citenamefont{Liu and Prokopec}(2017)}]{Liu_2017}
\bibinfo{author}{\bibfnamefont{L.}~\bibnamefont{Liu}} \bibnamefont{and}
  \bibinfo{author}{\bibfnamefont{L.}~\bibnamefont{Prokopec}},
  \bibinfo{journal}{Phys. Lett. B} \textbf{\bibinfo{volume}{769}},
  \bibinfo{pages}{281} (\bibinfo{year}{2017}).

\bibitem[{\citenamefont{Kardashev et~al.}(2013)\citenamefont{Kardashev,
  Khartov, Abramov, Avdeev et~al.}}]{Kardashev2013}
\bibinfo{author}{\bibfnamefont{N.~S.} \bibnamefont{Kardashev}},
  \bibinfo{author}{\bibfnamefont{V.~V.} \bibnamefont{Khartov}},
  \bibinfo{author}{\bibfnamefont{V.~V.} \bibnamefont{Abramov}},
  \bibinfo{author}{\bibfnamefont{V.~Y.} \bibnamefont{Avdeev}},
  \bibnamefont{et~al.}, \bibinfo{journal}{Astron. Rep.}
  \textbf{\bibinfo{volume}{57}}, \bibinfo{pages}{153–194}
  (\bibinfo{year}{2013}).

\end{thebibliography}
\bibliographystyle{apsrev}
\end{document}